\documentclass[reprint,showpacs,footinbib,citeautoscript,floatfix,nofootinbib]{revtex4-1}

\usepackage{graphicx}
\usepackage{amsmath}
\usepackage{amssymb}
\usepackage{mathrsfs}
\usepackage{bm}
\usepackage{array}
\usepackage{natbib}
\usepackage{soul}
\usepackage{natbib}

\DeclareUnicodeCharacter{2212}{-}
\newcolumntype {s}[1]{@{\hspace{#1}}} 
\usepackage{color}
\usepackage[usenames,dvipsnames]{xcolor}
\usepackage{wrapfig}
\usepackage[normalem]{ulem}

\begin{document}

\title{Rare-Earth Nitrides: Fundamental Advances and Applications in Cryogenic Electronics}

\author{W.~F.~Holmes-Hewett$^{1,2,*}$, J.~D.~Miller$^{1}$, H.~G.~Ahmad$^{3}$  S.~Granville$^{1,2}$ and B.~J.Ruck$^{2,4}$}

\affiliation{$^{1}$Robinson Research Institute, Victoria University of Wellington, P.O. Box 33436, Petone 5046, New Zealand}

\affiliation{$^2$MacDiarmid Institute for Advanced Materials and Nanotechnology, P.O. Box 600, Wellington 6140, New Zealand}

\affiliation{$^3$Dipartimento di Fisica Ettore Pancini, Università degli Studi di Napoli Federico II, c/o Complesso Monte Sant’Angelo, via Cinthia, I-80126 Napoli, Italy}

\affiliation{$^4$School of Chemical and Physical Sciences, Victoria University of Wellington, P.O. Box 600, Wellington 6140, New Zealand}

\affiliation{*Corresponding author: W.~F.~Holmes-Hewett, William.Holmes-Hewett@vuw.ac.nz}

\date{\today}

\begin{abstract}

Driven by the pursuit of high-performance electronic devices, research into novel materials with properties appropriate for cryogenic applications has unveiled the exceptional properties of the rare-earth nitride series of intrinsic ferromagnetic semiconductors. Here we report on the field focusing on developments, since the most recent comprehensive review~\cite{Natali2013}, which enable applications in cryogenic electronic devices. 

\end{abstract}

\maketitle

\section{Introduction}

\label{Background}

The energy cost of computing has skyrocketed over the last ten years, driven by the demands placed on the grid by the uptake of new technologies~\cite{Andrae2015, Jones2018, Masanet2020, Mehonic2022}. This increasing demand has proven an effective driver for the study and development of new, more energy-efficient devices, and research into the materials on which these devices rely. Spintronic devices offer lower power and faster operations than traditional CMOS platforms~\cite{Hirohata2020} and even greater efficiencies and speeds can be obtained where spintronic devices are paired with superconducting materials (the field of \textit{superconducting} spintronics)~\cite{Mukhanov2011,Tolpygo2007,Linder2015}. These devices have exciting applications in both low-energy superconducting classical and quantum computing where they have been imagined to form both the logic and memory systems. While there have been many recent advances in this area~\cite{Glick2017,Vernik2013,Baek2014,Nevirkovets2018,Nevirkovets2023, Rowlands2019, Nguyen2020}, the need for versatile materials to satisfy the requirements of next generation of spintronics and computing is becoming more pressing~\cite{Alam2023,Soloviev2017}. This review focuses on a single series of materials, the rare-earth mononitrides (RN - R a lanthanide), in this context. 

The tuneable electronic and magnetic properties of the RN are valuable from an applied perspective as their control can be leveraged in next generation spintronic and superconducting-spintronic devices. In addition, the series offers a unique platform to investigate the rich fundamental physics of the strongly spin-orbit coupled lanthanide ions. We begin with a historical introduction of the RN, then move to review recent progress in thin film formation and new understanding of the reaction between the lanthanide ions and molecular N$_2$ which facilitates the formation of the RN. We then focus on the materials which have seen the most attention in recent decades, GdN, SmN, and their alloys. We first discuss the magnetic properties, then the electronic. There will be a brief tour of the remaining members of the RN series and general notes of interest covering advances since the most recent comprehensive review~\cite{Natali2013}.

Finally, we describe the application of these materials in cryogenic electronics and their potential for superconducting spintronic devices. In conclusion we summarise the state-of-the-art of RN research, with some discussion of the outstanding questions in the field of these materials which would benefit from direct attention.

\section{Background}

The lanthanide elements, La (\textit{Z}=57) to Lu (\textit{Z}=71), are characterised by the varying occupation of their 4\textit{f} orbitals. They have electron configuration [Xe]6s$^2$4f$^\textit{n}$ (\textit{n}$=[0,14]$) while some also have a 5\textit{d} electron when this is more energetically favourable to an additional 4\textit{f}. In their elemental form they are metallic, with 4\textit{f} bands and 5\textit{d} bands crossing the Fermi level, while the reduced radius of the 4\textit{f} wave-function limits orbital quenching effects common in the transition metals. This leads the lanthanides to have strong orbital moments which combine with the spin to yield rich magnetic character. 

The mononitrides are among the simplest lanthanide compounds and to our knowledge were first reported on in 1902~\cite{Muthmann1902}. The early papers in the 1950s focused on the structural properties~\cite{Young1952,Klemm1956,Kempter1957} reporting the lattice parameters and rock-salt structure of several members of the series. The first descriptions of the electronic properties appeared in the early 1960s where most were reported to be semi-metallic and trivalent with the exception of SmN, YbN and EuN, which showed evidence of divalent lanthande ions, and CeN where Ce adopts a $4^+$ charge state~\cite{Didchenko1963}. Early optical studies found DyN, ErN, and HoN to be degenerately doped semiconductors, which was attributed to a deficiency of N~\cite{Sclar1964}. The first reports of the magnetism were via neutron diffraction where most of the RN were found to show a ferromagnetic order at cryogenic temperatures~\cite{Child1963,Trammell1963,Busch1965}.

The 1970s through 1990s saw reduced activity regarding the RN, although notably 1977 saw the first explicit report of the electronic structure of any of the series, with a computational study of Gd-pnictides~\cite{Hasegawa1977}. Of these all were metallic apart from GdN, notably the only member to show separated conduction and valence bands. The crystals produced during these decades suffered from oxygen contamination up to a few~\%, and the dependence of various properties with substitutional oxygen was studied in a few cases~\cite{Kuznietz1971,Gambino1970,Wachter1980,Brown1974,Li1991,Kasuya1997,Li1997}. Experimental reports focused on the magnetic properties of GdN~\cite{Gambino1970,Kuznietz1971,Cutler1975,Wachter1980}, SmN~\cite{Moon1979}, and YbN~\cite{Zhou1991,Bras1993,Kasuya1997c}. In addition there are some notable reviews of rare-earth pnictides~\cite{Hulliger1979,Vogt1993} with significant discussion of the RN.

Reference~\cite{Schneemeyer1987} is the first report (to our knowledge) of any of the RN grown as thin films, in this case GdN and a GdYN alloy deposited via sputtering, which would become a standard technique in the decades to come.  Another significant study in this period is the first report of a sputtered RN film interfaced with a superconductor, in this case GdN/NbN multilayers~\cite{Xiao1996}. 

In the following sections we will discuss the advances in thin film deposition techniques regarding the RN, the understanding of the reaction between the lanthanide metals and molecular nitrogen gas, the magnetic and electronic properties of the RN and in context explain why these materials are attractive for cryogenic electronics. We will then briefly review the use of these materials, largely GdN, in cryogenic and superconducting devices. 

\section{Growth and Formation}

\subsection{Thin Film Growth}
\label{growth}

The increasing availability of high-vacuum thin film growth capabilities in the 1980s and 90s, along with the combination of more powerful computer systems and advances in computational techniques, led to a significant resurgence of interest in the RN in the early 2000s. Modern thin film deposition techniques have allowed the production of samples with a much lower number of defects and impurities than the methods of the 1950s-1980s. There are now many reports of RN films grown by reactive magnetron sputtering~\cite{Leuenberger2005,Shalaan2006,Yoshitomi2011,Senapati2011a,Osgood1998,Shimokawa2015,Meeuwissen2021} pulsed laser deposition~\cite{Ludbrook2009,Tanaka2024}, chemical vapour deposition~\cite{Cwik2017,Brewer2010,Gernhart2014}, and physical vapour transport~\cite{Atabi2018}, yet molecular beam epitaxy~\cite{Melendez-Sans2024,Pereira2023,Anton2023,McNulty2021,Maity2020,Chan2016,Lee2015,Natali2014,Natali2012,natali2010,Gerlach2007} has proven the method of choice for films of the highest crystal quality. Here the orientation of the RN thin film can be dictated by the substrate choice, resulting in well ordered thin films of a single out of plane orientation. The push for epitaxial films of the RN has been so far motivated by fundamental studies. Here, the motivation is largely to understand the electronic structure of the material and ultimately the electronic ground state. This is somewhat complicated by the requirement of elevated temperature to facilitate epitaxial growth and the low activation energy of nitrogen vacancies~\cite{Punya2011,Porat2024}, resulting in heavy electron doping for most of the epitaxial films reported. 

For most device structures investigated to date, epitaxial growth has not been possible. The limited range of lattice matched materials, in combination with the requirement of capping layers to protect the air sensitive RN, limits both the range of potential substrates and traditional post-processing techniques. Thus, the RN films discussed in the context of device structures are mostly polycrystalline. These, in general, have poorer crystal quality (leading to a decreased mobility) and lower carrier density (stemming from a lower population of nitrogen vacancies) leading to more resistive films.

Many of the reports in the early 2000s on polycrystalline films (see Ref.~\cite{Natali2013} for a review), were sufficient to reinvestigate the magnetic and electronic ground states of the RNs, first prepared decades earlier. Yet, while studies of the structure of the nitrides have largely conformed to the picture developed in the last century, a study on polycrystalline thin films of GdN deposited on amorphous fused silica has investigated the possibility of mixed structural phases forming due to nitrogen vacancies~\cite{Shaib2020}. This paper follows earlier reports opening the possibility of two distinct structural phases~\cite{Senapati2011a,Senapati2010}, but these papers do not agree on the magnetics of the additional structural phases proposed. 

The first reports of epitaxial RN thin films arise in 2003~\cite{Lee2003} when CeN thin films were grown on MgO via reactive magnetron sputtering. MgO is a common substrate and shares the cubic rock-salt structure with the RN. However, the significant ($\approx~15$~\%) lattice mismatch is near the largest possible that is not prohibitive for epitaxial growth. In addition to CeN, only GdN~\cite{Gerlach2007,Tanaka2024}, SmN~\cite{Anton2013,Vallejo2024} and YbN~\cite{Loyal2023} have been grown epitaxially on MgO. Moving to aother substrates, the RN materials form well-ordered [111] orientated structures on the [0001] hexagonal net of AlN or GaN~\cite{SCARPULLA2009,Yoshitomi2011,Vezian2016,Chan2017}. More detailed studies~\cite{Chan2016} of the growth of GdN and SmN on the Si:AlN surface discuss the formation of multiple rotational domains, and the transition to a [001] dominated growth mode at elevated temperatures. The existence of multiple rotational domains in otherwise epitaxial films is implicated in the appearance of perpendicular magnetic anisotropy (PMA) in these samples~\cite{Miller2023}. Epitaxial films have also been achieved by deposition of 6~nm layers of Gd or Sm on AlN and introducing nitrogen gas~\cite{Chan2023}. Other substrates investigated include YSZ on which EuN~\cite{Richter2011,Ruck2011}, GdN~\cite{Ludbrook2009,Natali2012} and SmN~\cite{Natali2012} have been studied. YSZ provides a good lattice match, but the mobile oxygen ions present in the structure result in an oxide of some nm thick at the substrate interface, limiting the effectiveness of YSZ as a substrate.

More recently there has been effort put into forming well ordered single domain thin films, requiring a shift from the Si:AlN system and the resulting twinned 111 rotational variants~\cite{Chan2016}. The lattice of the RN (experimental lattice parameters spanning 4.76~\AA to 5.305~\AA~\cite{Natali2013}) match reasonably well to Si (5.43~\AA). However, an interfacial reaction forming RSi$_2$~\cite{Youn1988,Natali2011} prohibits the use of bare Si without extreme care. A method to deposit RN films, specifically SmN, directly onto bare Si was demonstrated~\cite{McNulty2021}, resulting in well-ordered [001]-orientated films. The symmetry of the cubic-on-cube [001] growth facilitates a single in-plane rotational domain, contrasting the two ([111] orientated RN) and three ([001] orientated RN) domain structures present in the AlN:Si systems. Following the initial studies on Si substrates there have been several reports studying the growth of RN materials on aluminate substrates. LaAlO$_3$ has been used for the growth of HoN~\cite{Pereira2023}, SmN~\cite{Melendez-Sans2024,Anton2023}, DyN~\cite{Anton2023}, GdN~\cite{Anton2023,Trewick2025}, while YAlO$_3$ has been used for LuN~\cite{Guanhua2024} and GdN~\cite{Tanaka2024}. These aluminate substrates have structures well suited to the growth of the RN. LaAlO$_3$ is pseudo-cubic with a lattice parameter of 3.78~\AA, this is $\approx 1/\sqrt{2} \times~5$~\AA~resulting in a 45$^\circ$ rotated cube-on-cube epitaxy. YAlO$_3$ is orthorhombic, and along the [110] face has a similar relationship resulting again in a rotated cube-on-cube epitaxy. These substrates allow the formation of well ordered films with a single in-plane rotational domain, without the issues of oxide formation of YSZ, or the reaction with Si.

Along with the interest in the structure and growth of the RN there have been recent reports of their vibrational properties. The optically active TO($\Gamma$) mode vibrational frequencies have been reported for GdN~\cite{holmes2019}, SmN~\cite{holmes2019}, NdN~\cite{Holmes-Hewett2019a}, DyN~\cite{Holmes-Hewett2020}, YbN~\cite{Holmes-Hewett2022} and LuN~\cite{Holmes-Hewett2022}. Reference~\cite{Holmes-Hewett2022} also reports the damping of the TO($\Gamma)$ vibration. Computational studies of the vibrational structure have been undertaken for ErN~\cite{Upadhya2022a}, LuN~\cite{vanKoughnet2023,Li2017} and GdN~\cite{vanKoughnet2023}. An early study~\cite{Granville2009} attributed the presence of a Raman signal in the RN to defects in the form of nitrogen vacancies. A recent Raman and computational phonon-dispersion study of GdN and LuN demonstrates that the strongest LO($\Gamma$) feature is intrinsic, activated by the Fr{\"o}hlich interaction~\cite{vanKoughnet2023}. 

\subsection{Catalytic Reaction}

As is commonly reported in the experimental literature RN thin films are formed by evaporating metallic lanthanide charges in the presence of molecular nitrogen gas (N$_2$), resulting in the formation of RN, the mononitride~\cite{Natali2013}. Other precursors have been used, notably NH$_3$~\cite{Chan2016,SCARPULLA2009}, and activated N$_2$ via a Kaufman ion or a plasma source~\cite{SCARPULLA2009}. Formation of the mononitride with N$_2$ implies that contact with the clean lanthanide surface facilitates the breaking of the molecular nitrogen bond at room temperature and low pressure. Breaking this bond is a process which typically requires temperatures and pressures elevated significantly over ambient conditions (400~$^\circ$C, 200~bar). From a commercial and industrial perspective, a lower energy method to cleave the nitrogen bond is attractive, for example, in the production of ammonia (NH$_3$). The industry standard Haber-Bosch process used to facilitate ammonia production uses several percent of the world's total energy budget, and in this context, a lower-energy alternative is attractive. 

\begin{figure}
\centering
\includegraphics[width=\linewidth]{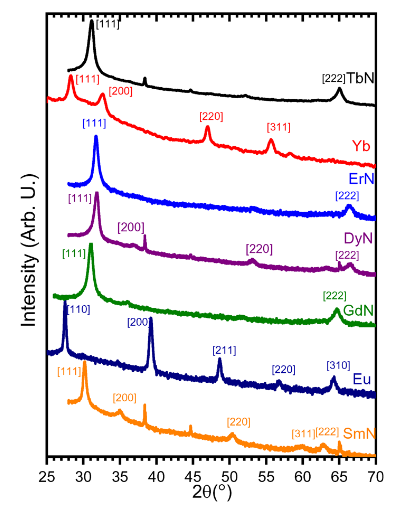}
\caption{X-ray diffraction measurements on various lanthanide metals deposited in a UHV system in a N$_2$ atmosphere at ambient temperature. Most lanthanide metals spontaneously form the mono-nitride under these conditions, with the exception of Eu and Yb. Figure reproduced from Reference~\cite{Ullstad2019}.}
\label{XRD}
\end{figure}

Reference~\cite{Ullstad2019} studies the catalytic reaction between the lanthanide metals and N$_2$ via the growth of RN films under a range of conditions, including the exposure of recently deposited lanthanide metals to molecular nitrogen. All of the lanthanide metals studied, apart from Eu and Yb, the two which favour a divalent state~\cite{Aerts2004}, yielded the formation of the RN upon exposure to molecular nitrogen gas. Figure~\ref{XRD} shows X-ray diffraction measuremnts of various lanthanide metals evaporated in a molecular nitrogen atmosphere. The process was primarily studied via \textit{in-situ} resistance measurements relying on the significant contrast between the electronic properties of the metallic lanthanide metals and the insulating RN. The reaction was found to be initially rapid implying a near instant reaction for the first few mono-layers. The initial reaction is followed by a slower diffusion-limited process implying that subsequent reaction is facilitated by transport of N or N$_2$ through the first few layers of recently formed RN. In addition, Reference~\cite{Ullstad2019} also finds that when an as-grown, or fully reacted, thin film of RN is exposed to vacuum, nitrogen is desorbed from the first few nm of the material.  

A following study~\cite{Chan2020} focusing on the facile breaking of the molecular nitrogen bond at a clean lanthanide surface, found that the reaction catalysis discussed in Reference~\cite{Ullstad2019} is significantly inhibited by the presence of a small partial pressure of oxygen, indicating the presence of a self-limiting oxide. In this report, Gd metal was grown on AlN then exposed to molecular nitrogen, with the exposed \textit{c}-plane of the Gd metal monitored by RHEED. The resulting RN was found to be of high structural quality, implying that passive nitridation of metallic lanthanide surfaces may be a fruitful avenue for the formation of high-quality thin layers of RN that are appropriate for vertical transport spintronic devices.

\begin{figure}
\centering
\includegraphics[width=\linewidth]{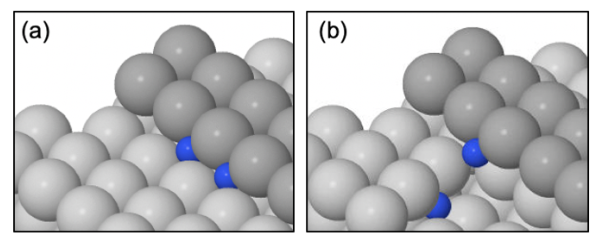}
\caption{Selected binding configurations of 2N on stepped, close-packed lanthanide metal surfaces. Step atoms are shown in darker gray. (a) Both N adsorbed in the step feature (“step-step”). (b) One N atom in the step feature and the other in the terrace subsurface layer (“step-terrace”). Note that R atoms in (b) have been removed for ease of viewing. Figure has been reproduced with permission from Reference~\cite{Chan2020}.}
\label{binding}
\end{figure}

Reference~\cite{Chan2020} also details a molecular dynamics study of the incorporation of a nitrogen molecule into a lanthanide surface. It reports that, similar to transition metal systems, the active absorption site is a monatomic step on a close-packed lanthanide surface. However, the geometry of the absorption differed significantly from the transition metal systems. The nitrogen molecule is absorbed \textit{into} the surface at the step edge, rather than \textit{onto} the surface; the two such binding configurations discussed in Reference~\cite{Chan2020} are shown in Figure~\ref{binding}. Of the systems studied (Nd, Sm, Eu, Yb) they find that the activation energy was lower for Nd and Sm than Eu and Yb, consistent with the experimental finding that the latter two lanthanides do not form the RN without the presence of activated nitrogen. Reference~\cite{Chan2020} finally demonstrates directly the formation of NH$_3$ using lanthanide metals, N$_2$, and hydrogen as reactants. A newly deposited lanthanide metal was first exposed to N$_2$ and then H$_2$. Upon the exposure to H$_2$ the presence of NH$_3$ was detected in the growth chamber, demonstrating (i) breaking of the N$_2$ and (ii) the formation of ammonia.

\section{Magnetics}

\subsection{Magnetic Properties of Single Cation RN}

\begin{figure}
\centering
\includegraphics[width=\linewidth]{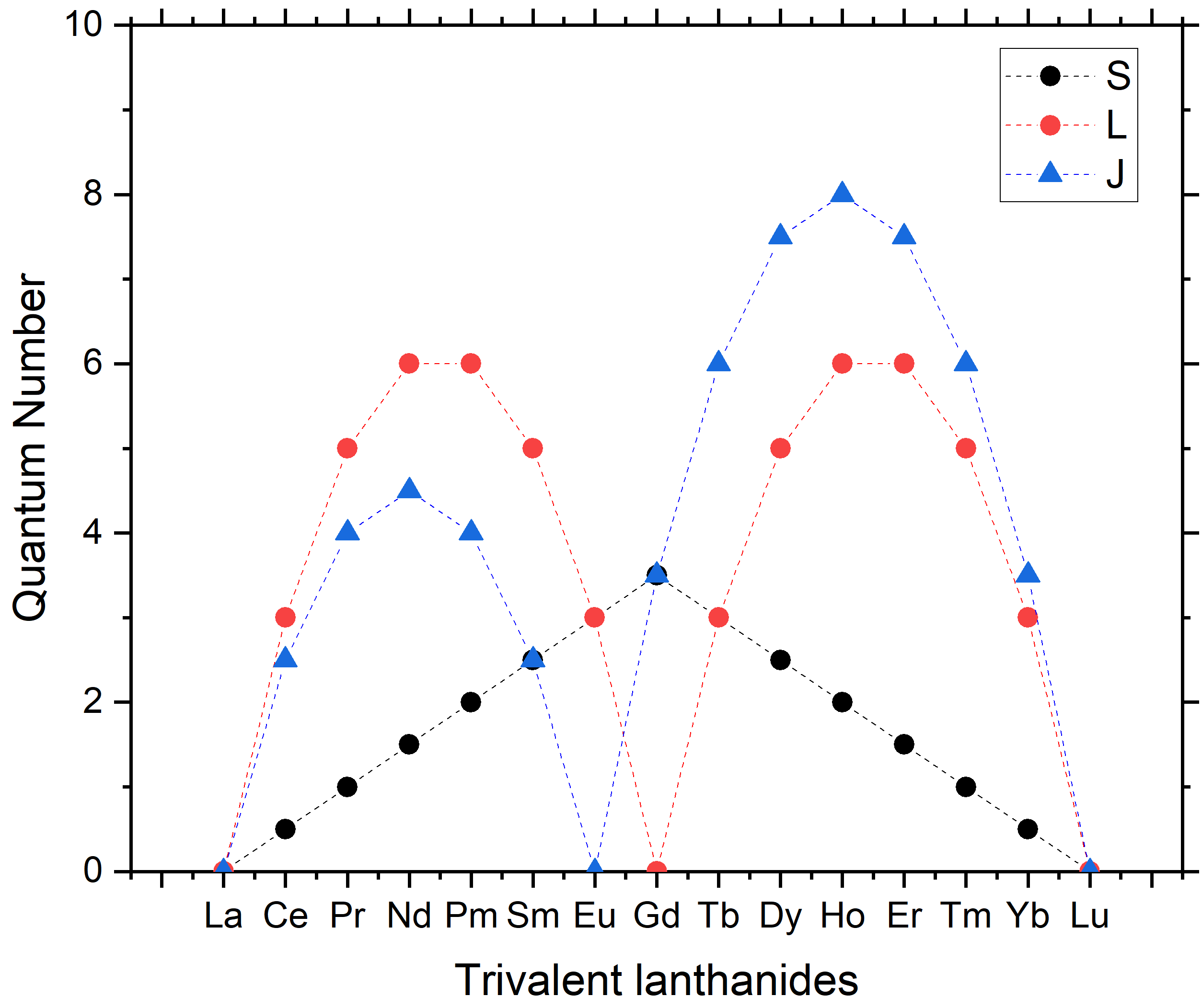}
\caption{Total angular momentum (\textit{J}), Orbital angular momentum (\textit{L}) and spin angular momentum (\textit{S}) for the trivalent lanthanide ions as determined from Hund's rules.}
\label{Quantum-numbers}
\end{figure}

In contrast to the transition metals, the lanthanide ions display a large unquenched orbital contribution to the angular momentum. This is a result of the reduced extent of the localized 4\textit{f} wave-functions. In the lanthanides these see only limited effects of the crystal field, which is the origin of the orbital quenching effects in the spatially extended \textit{d} wave-functions of the transition metals. In lanthanide metals this manifests as a total angular momentum formed from a combination of, or competition between, the spin and orbital parts. Although far from complete, Hund's rules provide a simple manner to evaluate the ground state quantum numbers for a given electronic configuration of an isolated ion. 

Figure~\ref{Quantum-numbers} shows the magnitude of the quantum numbers \textit{S}, \textit{L} and \textit{J} for the trivalent lanthanide ions, based on Hund's rules. Here, \textit{J} is the total angular momentum quantum number associated with the ground state \textit{J}-multiplet given by Hund's third rule. For the light rare-earths (4\textit{f} shell is less than half full), the spin-orbit interaction dictates that the ground state $J=\lvert L-S \rvert$, while for the heavy rare-earths (4\textit{f} shell is more than half full) the lowest energy  \textit{J}-multiplet is instead given by their sum $J=L+S$~\cite{Ashcroft1976}.

\begin{figure}
\centering
\includegraphics[width=\linewidth]{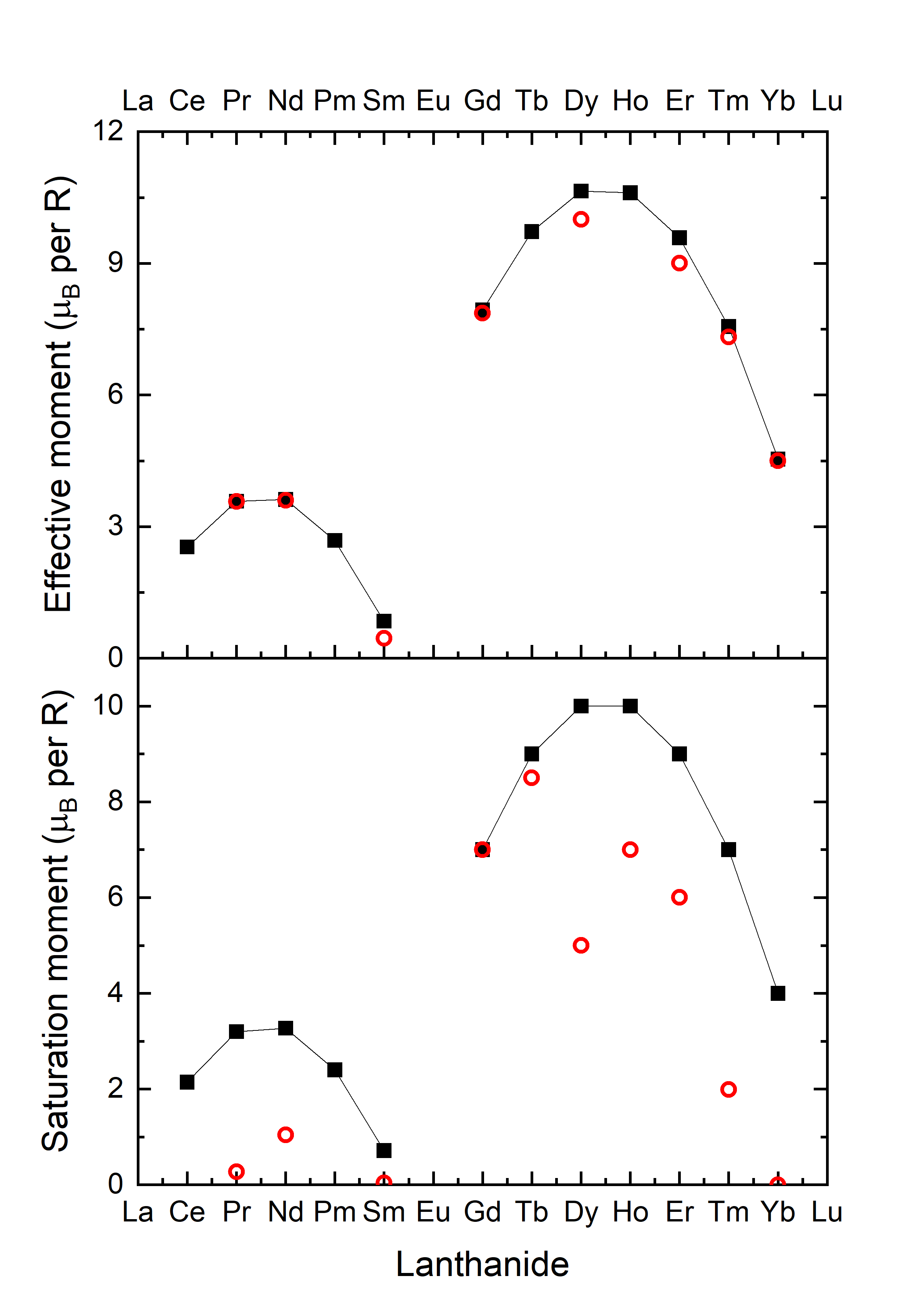}
\caption{Effective moments for trivalent rare-earth ions based on Hund’s rules for \textit{J}, \textit{L} and \textit{S} (black squares) and experimental values (red circles), for various RN, of (a) the effective moments measured in the paramagnetic phase and (b) the saturation moments in the ferromagnetic phase.}
\label{Moments}
\end{figure}

In the free-ion Hund's rules ground state description the value of the effective moment for the trivalent lanthanide ions is given by 

$$\mu_{\mathrm{eff}}=g_J\mu_B\sqrt{J(J+1)},$$

\noindent with $g_J$ the Landé g-factor. These are illustrated in panel~(a) of Figure~\ref{Moments}, which shows that the lanthanides have effective moments ranging from 10.63~$\mu_B$ in Dy to 0.84~$\mu_B$ in Sm. Also shown are the reported values measured in various RN. As is common in lanthanide compounds the effective moment of the RN is consistent with the free-ion Hund's rules moment in the paramagnetic phase well above the Curie temperature (ErN~\cite{Meyer2010}, GdN~\cite{natali2010}, YbN~\cite{Warring2014}, NdN~\cite{Anton2016b}, DyN~\cite{Shaib2021}). However, the admixture of higher energy \textit{J} multiplets does affect the simple picture in some cases (SmN~\cite{Meyer2008}, EuN~\cite{Ruck2011,Binh2013}, NdN~\cite{Anton2016b}) where there is a complex multiplet structure at accessible energy levels.  

The measured ferromagnetic moments in the nitrides show more deviation from the Hund's rules expectation. As shown in panel~(b) of Figure~\ref{Moments}, in all members apart from GdN, and the trivial cases of LaN and LuN (not shown), the saturation moments in the ferromagnetic phase are significantly less than the simple expectation

$$M_{S}=g_JJ.$$

\noindent The reduction is a consequence of the combined influence of the crystal field and the exchange interaction~\cite{Meyer2010, Anton2016b, McNulty2016,Shaib2021,Evans2017}. 

The simplest ferromagnetic member of the series, GdN, has a half-filled 4\textit{f} shell, yielding zero orbital angular momentum $(L=0)$ and thus a spin-only magnetic moment $(J=S)$, and a multiplet structure beyond a typical energy scale. The large GdN saturation moment $M_{S}=7\mu_B$ per Gd ion, which matches the Hund's rules expectation, brings in concert a strong Zeeman interaction with an applied field, and thus a small coercivity on the order of 100~Oe or less~\cite{natali2010,Ludbrook2009}.

This apparent simplicity of the magnetic state has resulted in GdN being the subject of significantly more studies than the other members in the series. A complication is that experimental thin film Curie temperature values vary - most reports fall into the range of 60-70~K ~\cite{Leuenberger2005,Granville2006,Senapati2010,Vidyasagar2014,Alfaro2019,Tanaka2024} but a number of studies on films made by magnetron sputtering report Curie temperatures approximately 30-40~K~\cite{Yoshitomi2011,Mi2013,Shimokawa2015,Vidyasagar2017}. While not fully resolved, the cause is considered to be differences in stoichiometry, strain, or grain size. It is clear that the presence of nitrogen vacancies can significantly affect the magnetic properties. References~\cite{Natali2013a,Shaib2020} demonstrated a double magnetic transition in films grown by molecular beam epitaxy with a controlled concentration of vacancies. Experiments show that films with a low density of nitrogen vacancies have T$_C\approx$50~K, whereas those with a higher concentration have T$_C\approx$70~K~\cite{Natali2013a}. Films with even greater nitrogen deficiency have shown higher T$_C$ values, even reaching above 150~K~\cite{Plank2011}.

The pairing mechanism in the ferromagnetic phase of GdN is not fully understood. While there are clear similarities to EuO, both with electronic configuration 4\textit{f}$^7$ and most commonly reported Curie temperature of $\sim$~70~K, the two do not appear to share the same pairing mechanism~\cite{Sharma2010}. Significantly, the 4\textit{f} states sit considerably lower in the valence band in GdN ruling out the pairing mechanism common in Europium chalcogenides~\cite{Sharma2010}. Pairing in GdN has been proposed to operate via a \textit{d-f} mixing and exchange interaction~\cite{Kasuya1997,Kasuya1997b}, with an indirect N~2\textit{p} mediated coupling~\cite{Mitra2008a}, further enhanced by a carrier-mediated model~\cite{Sharma2010}. In addition to the work on GdN there has been a detailed study of the multi-polar exchange interaction in NdN~\cite{Iwahara2022}. The consensus of these results is that there is some combination of indirect exchange present in these materials, although the details of this remain unclear.

Moving away from the simplicity of the 4\textit{f}$^7$ configuration of GdN, the orbital angular momentum, and its contribution to the magnetic moment, now become non-zero for the remaining RN, with the total angular momentum a combination of both \textit{S} and \textit{L}. SmN is likely the second most studied member of the series, and in this regard provides an interesting contrast to GdN. The 4\textit{f}$^5$ configuration of the free Sm$^{3+}$ ion has $S=5/2, L=5, J=5/2$, with \textit{S} and \textit{L} opposing in the less than half filled 4\textit{f} shell. In the ferromagnetic phase of SmN, the influence of the crystal field and exchange results in a near complete compensation between the opposing spin and orbital contributions to the magnetism. This results in full spin alignment, both intra- and inter-ion, yet a near zero net magnetic moment, experimentally measured as 0.035~$\mu_B$ per Sm ion~\cite{Meyer2008} with the residual orbital moment directed anti-parallel to the spin~\cite{Anton2013}. The near-zero moment of SmN results in a much weaker Zeeman interaction, compared to GdN, and a coercivity orders of magnitude higher, greater than 9~T at 2~K~\cite{Meyer2008,Holmes-Hewett2018}. The common structure and growth modes of GdN and SmN in combination with the significantly contrasting magnetic properties make them an interesting hard-soft magnetic pair. 

The near-zero moment of SmN is not unique in ferromagnetic materials. Particularly, reports exist of ferromagnetic (Gd,Sm)Al$_2$, an \textit{identically}-zero moment ferromagnetic material~\cite{Adachi1999,Adachi2001,Qiao2004}. However, for this material the zero moment compensation point depends critically on composition, temperature and field history~\cite{Taylor2022,Avisou2007}. Contrasting (Gd,Sm)Al$_2$, the compensation in SmN is centred entirely on the Sm ions, it is robust with temperature below the Curie temperature of 30~K, and due to the large coercivity is mostly not affected by an applied field~\cite{Meyer2008}. Significantly, this compensation is also present at the technologically important temperature of 4~K, the operating point of many superconducting electronic technologies~\cite{Alam2023}. 

Other members of the RN have seen interest regarding their magnetic behaviours, although less so than GdN and SmN. DyN~\cite{Cortie2014,Evans2017, Shaib2021} and ErN~\cite{Meyer2010} have 4\textit{f} occupations greater than seven, and thus the spin orbit interaction dictates parallel spin and orbital angular momentum, as described above. These materials are ferromagnetic with saturation moments reduced from the simple expectation $g_JJ$, described in some detail in References~\cite{Cortie2014,Evans2017,Shaib2021} and~\cite{Meyer2010} for DyN and ErN, respectively. For the lighter members, which have orbital dominated magnetic moment, only NdN has seen recent interest~\cite{Anton2016b,Holmes-Hewett2019a}, again with a magnetic moment reduced from simple expectation. YbN is relatively anomalous, as it the only known member to display anti-ferromagnetic coupling~\cite{Degiorgi1990,Warring2014}. Finally, EuN~\cite{Ruck2011,Binh2013} provides an interesting case. As is common in trivalent Eu compounds, there is a complete compensation between the spin $(S=3)$ and orbital $(L=3)$ parts of the angular momentum on the Eu ions. The $J=0$ state does not, in principle, allow alignment. However, as described in References~\cite{Ruck2011,Binh2013} a small concentration of strongly magnetic divalent Eu, a result of nitrogen vacancies, facilitates alignment of the Eu$^{3+}$ ions by admixture of the $J=1$ multiplet, courtesy of the inter-ion exchange. The resulting picture is that EuN is an intrinsic dilute magnetic semiconductor, dopable with nitrogen vacancies rather than magnetic ions~\cite{Binh2013}. 

\subsection{Magnetic Properties of Mixed Cation (R,R')N}
\label{S:Magnetic:Alloys}

\begin{figure}
\centering
\includegraphics[width=\linewidth]{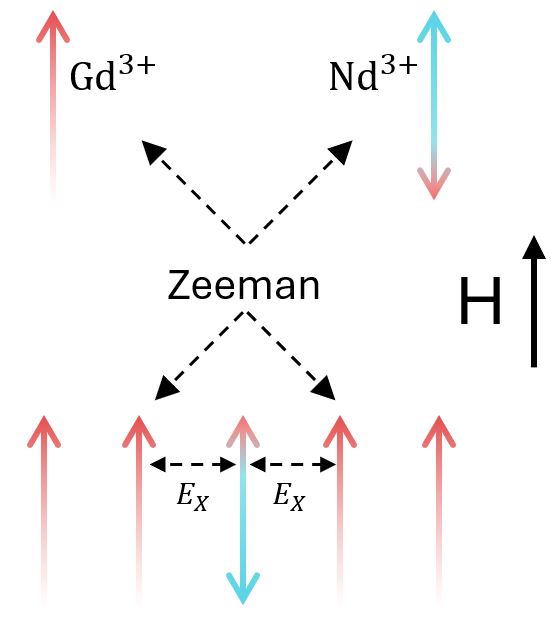}
\caption{Schematic representation of the competition between Zeeman and exchange coupling in the RN, showing both spin (red arrows) and orbital (blue arrows) contributions to the magnetic moments. The Gd$^{3+}$ and Nd$^{3+}$ ions of GdN and NdN align via the Zeeman interaction with their net magnetic moments parallel to an applied field. Note that in NdN this results in the spin magnetic moment being anti-aligned to the applied field. For the case of Nd$^{3+}$ ions dilutely doped into a GdN matrix, at low fields the inter-ion Gd-Nd exchange dominates the Zeeman interaction on the Nd$^{3+}$ ion, aligning the spin with the Gd$^{3+}$, and the magnetic moment of the Nd$^{3+}$ is anti-parallel to the applied field.}
\label{Zeeman}
\end{figure}

The selection of R in the RN provides a wide variety of magnetic materials as discussed above. Even greater control is found by forming alloys of various R in the nitride (i.e., (R,R')N). The RN provides a range of miscible materials where, to first order, the magnetic properties can be interpolated between the end members R and R' in binary alloys. One theoretical study exists of (La,Gd)N, which shows the lattice parameter varies continuously between LaN and GdN, with only a small variation from a Vegard's law linear dependence on R concentration~\cite{Mahfoud2018}, although we note that, as described in Section~\ref{growth}, the lattice parameter of a given RN can vary with preparation conditions. In all cases studied so far, the cation concentration appears to not alter the inter-ion ferromagnetic exchange in the nitrides, while the intra-ion spin-orbit physics is also retained. Note that it is the \textit{spin} states of the 4\textit{f} electrons on adjacent R ions which are coupled by the exchange interaction. In this way, if one considers a R ion from the first half of the series, say Nd$^{3+}$, which as a first approximation has \textit{L} and \textit{S} anti-parallel to each other, in a matrix of R ions from the second half of the series, such as Gd$^{3+}$ ions, as pictured in Figure~\ref{Zeeman}, the spin of the Nd ion will couple to the spin of the Gd ions. The \textit{L} of the Nd ion is anti-parallel to the coupled \textit{S} of the two ions. The Nd ion \textit{L} is larger than its \textit{S}, hence it follows that there will be angular momentum and magnetization compensation points as a function of cation concentration in this material. In the case of a dilute concentration of Nd in (Gd,Nd)N, if an external field is applied to the material, the strong Zeeman interaction on the more numerous Gd$^{3+}$ ions aligns the magnetic moment and spin angular momentum with the applied field. The exchange coupling then orients the spin of the Nd$^{3+}$ such that the spin moment is aligned with the applied field, and thus the total magnetic moment of the Nd$^{3+}$, which is dominated by the orbital contribution, aligns anti-parallel with the applied field. 

\begin{figure}
\centering
\includegraphics[width=\linewidth]{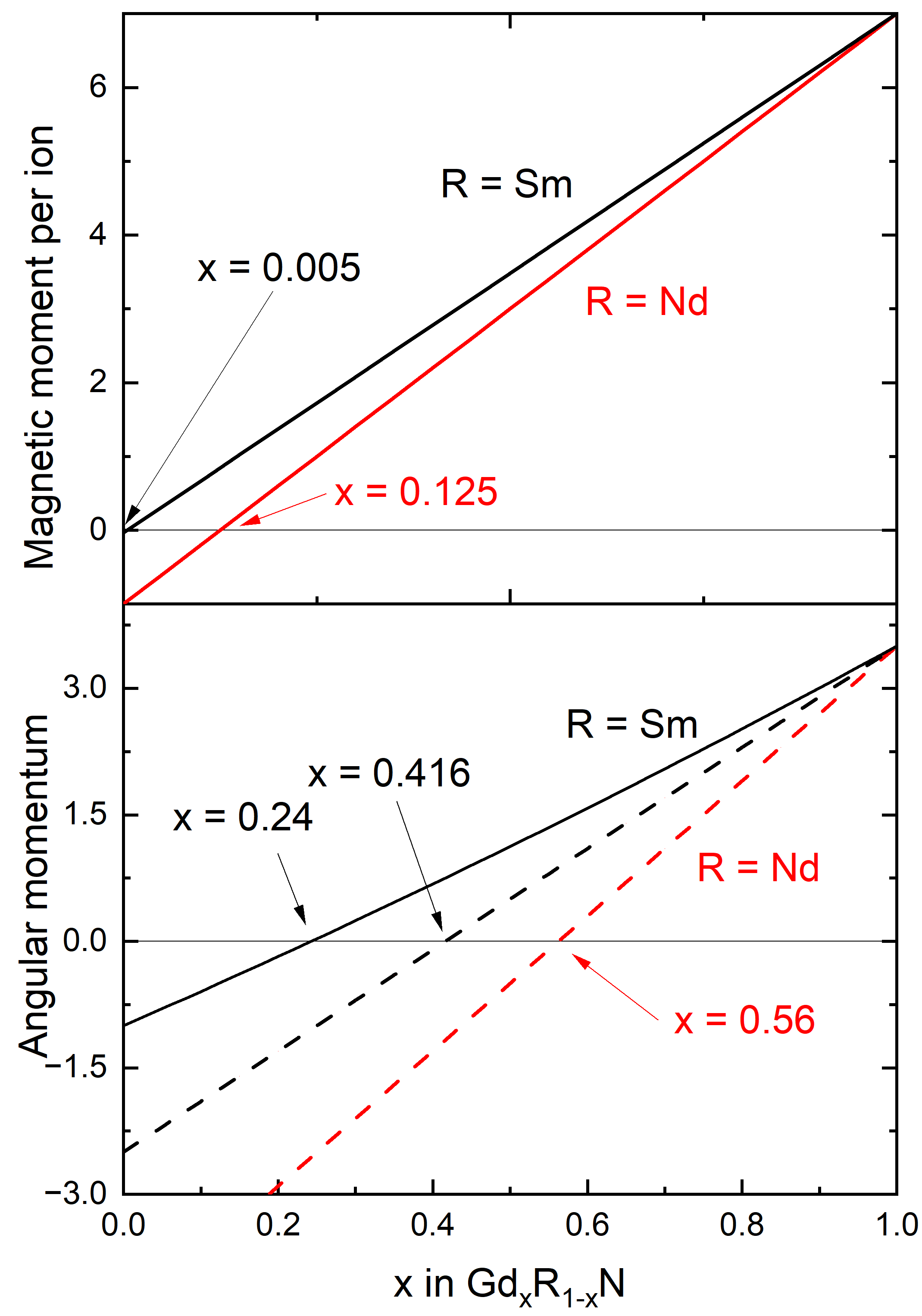}
\caption{Magnetization (a) and angular momentum (b) of Gd$_x$R$_{1-x}$N for R$=$(Sm,Nd). (a) Based on linear extrapolation of the experimental saturation magnetic moment of the end members. (b) Angular momentum extrapolated from the Hund's rules ground state of the end members (dashed lines), and informed by experiment (solid line)~\cite{McNulty2016,Miller2022}.}
\label{Compensation}
\end{figure}

These pairings of two lanthanides with competing spin and orbital-dominated angular momentum open the possibility of investigating various compensation points, possibly where the greatest technological potential is to be found. The $M=0$ compensation point may be used to form a perfectly hard ferromagnetic material, with no Zeeman coupling to an external magnetic field and zero fringe field, while the $J=0$ compensation point could facilitate efficient switching in spin-torque devices, as has already been investigated in compensated ferri-magnetic systems~\cite{Stanciu2006,Ueda2017,Siddiqui2018,Finley2019}. A first approximation of the compensation points can be found by linear extrapolation between the magnetic moments and angular momenta of the end members. Figure~\ref{Compensation} shows the examples of (Gd,Sm)N and (Gd,Nd)N with the magnetic moment (panel~(a)) and angular momentum (panel~(b)) determined by linear extrapolation of experimental measurements of the end members. 

Unsurprisingly, the most studied of these alloys is (Gd,Sm)N. Here the angular momentum, magnetization, and coercivity can be controlled, with the angular momentum running from the 3.5~$\hbar$ of GdN to -1~$\hbar$ in SmN, once accounting for the crystal field and exchange interaction in SmN~\cite{McNulty2016} and the enhanced spin projection of the Sm ion in SmN~\cite{Miller2022}. The zero moment and angular momentum points are indicated on Figure~\ref{Compensation}. The coercivity spans from some 100s of Oe in the Gd rich (Gd,Sm)N films to $>~5~T$ in Sm rich films~\cite{pot2023,Porat2024}, we note there is an expected divergence at the magnetic compensation point. Additionally Reference~\cite{Miller2023} notes the substitution of a small amount of Gd into SmN can produce a perpendicular magnetic anisotropy in thin films.

The only other binary alloy of the RN studied so far is (Gd,Dy)N, where again the magnetic properties roughly track between the end members~\cite{Pot2021}. Both Gd$^{3+}$ and Dy$^{3+}$ have net magnetic moments aligned with the spin contribution and thus there are no compensation points to investigate in this material. 

\subsection{Bi-layers and Super-lattices}
\label{S:Magnetic:Multi}

The RN provide a series of versatile materials with starkly contrasting magnetic properties, yet structurally they are very similar, with lattice parameters differing by $\sim$~10~\% between the end member extremes. As such, hetero-epitaxy is clearly possible and thus the magnetic properties of bi-layers, multi-layers and super-lattices have been the subject of some studies~\cite{McNulty2019,Anton2021,Natali2017,Alfaro2019,McNulty2015} due to the clear technological importance of these structures. As discussed above, coupling between neighbouring R ions is via the exchange interaction, determined by the spin of an ion rather than the net magnetization. This remains the case at RN:R'N interfaces and results in complex magnetic structures when these interfaces comprise materials of contrasting spin- and orbital-dominated magnetism, i.e., opposite spin orientations. Of particular interest here are exchange springs and engineered domain wall structures. When combined with the independently tunable electronic and magnetic properties of the RN, there is opportunity to probe the interplay between inhomogeneous magnetic order and spin-polarized transport.

\begin{figure*}
\centering
\includegraphics[width=\linewidth]{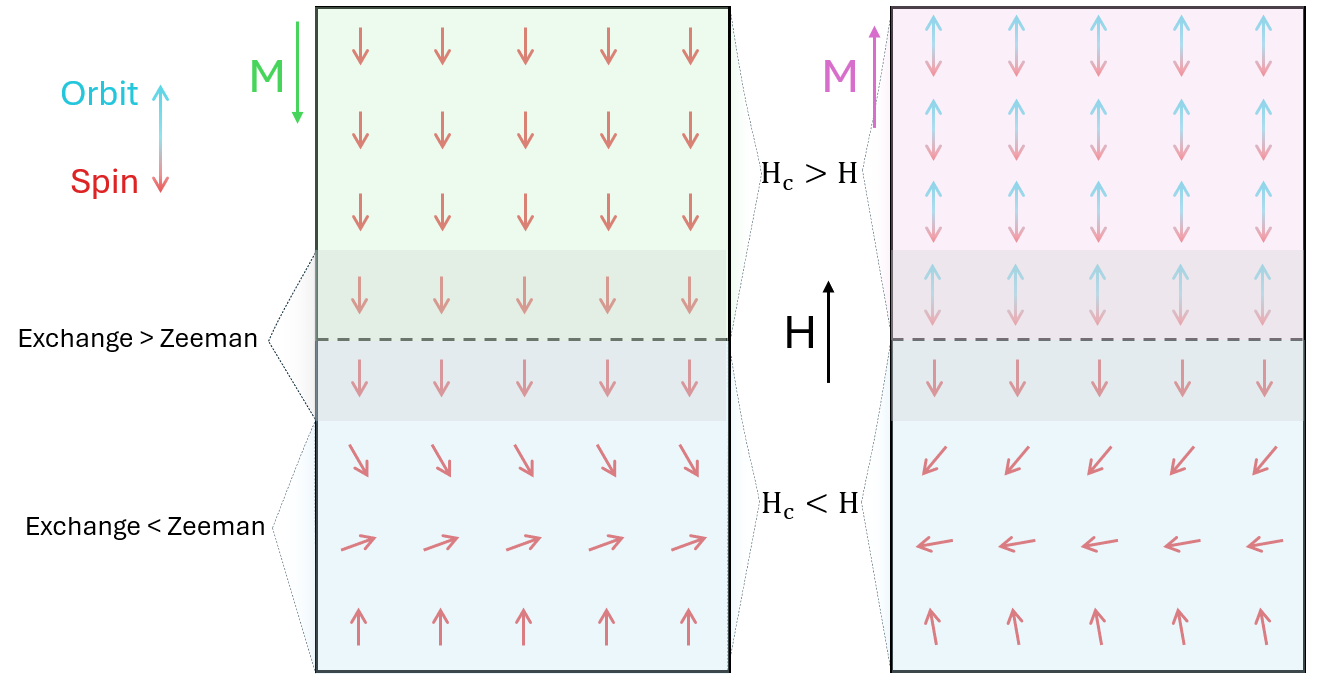}
\caption{Schematic of two exchange springs featuring hard ferromagnetic materials with spin (left) and orbital (right) dominated magnetic moments, both soft layers have spin dominated magnetic moments. In the left image the hard ferromagnetic layer is anti-allied to the applied field, resulting in a conventional exchange spring at the interface. In the right image the hard layer has an orbital dominated magnetism which is aligned to the applied field, resulting in a non-conventional exchange spring effect.}
\label{Spring}
\end{figure*}

A conventional exchange spring system comprises two spin-dominated ferromagnets (see Figure~\ref{Spring} left panel), typically one hard and one soft, with a shared interface, in which a modest field can rotate the alignment of the soft layer leaving the alignment of the hard layer unchanged~\cite{Fullerton1998}. In the anti-parallel alignment, exchange coupling at the interface pins the spins of the soft-layer to those of the hard layer. Deeper into the soft-layer the spins rotate towards the direction parallel to the applied field resulting in the exchange spring effect. Exchange springs have been demonstrated in the RN, by forming a bi-layer from a combination of light and heavy RN, where more complex behaviours can be seen. 

The case of bi-layers of GdN and SmN has been studied in Reference~\cite{McNulty2015}. Here GdN and SmN have spin and orbital dominated magnetic moments respectively (see Figure~\ref{Spring} right panel), with the spin component in SmN anti-aligning to the net magnetization. If such a bi-layer system is prepared in a conventional \textit{parallel} configuration of the magnetic moments in the two layers, the spins in the bulk are now anti-aligned. XMCD studies on these systems have shown that the spins on the Sm ions at the GdN:SmN interface couple to those of the GdN, a result of the combined strong inter-ion exchange and weak Zeeman interaction on the Sm ions. In the magnetically aligned state the spins form a conventional exchange spring, while the magnetizations form a more complex twisted structure. GdN:NdN layered structures have also been investigated~\cite{McNulty2019}. Here, there are particularly complex behaviours at the interface, relating to the competing exchange and Zeeman energy scales, resulting in a negative remanent magnetization. These engineered magnetic systems provide a rich playground for fundamental studies into magnetic configurations determined by competition between field- temperature- and thickness-dominated energy terms. The technological exploitation of these same systems is already being demonstrated as discussed in section~\ref{Applications} below. 

\section{Electronic Structure} 

It was largely the electronic structure of the RN series which resulted in the enhanced interest from the mid-2000s compared to the other R-pnictides. The bond is largely ionic between the trivalent R and nitrogen ions, which in the stoichiometric form find $3+$ and $3-$ charge states respectively. Although there are still debates about the nature of the electronic ground state of many of the members, the majority of recent experimental studies find that at least GdN is insulating when nominally stoichiometric. 

The first electronic studies on the RN were conducted on bulk samples in the 1950s, where they were thought to be semimetallic. With the increasing availability of computational systems in the 2000s and the associated advances in first-principles electronic structure calculations, interest in computational studies on these materials grew, with particular attention paid to their highly localised 4\textit{f} states. A systematic study of the full series of RN found several to have half-metallic ground states~\cite{Aerts2004}, and associated application in spintronic devices. Although it is now understood that this result was due to difficulty in treating the correlation on the lanthanide~\textit{f} and \textit{d} states, it served to reignite modern experimental interest in the RN. To complement the advances in computational methods, the improved preparation technology available by the 2000s made the study of high-quality thin films possible, so the ground states of the different RN could finally be reinvestigated. Unintended nitrogen vacancies~\cite{Granville2006}, acting as n-type donors, were identified as the reason for concluding semi-metallic ground states from the earlier experiential studies, at least for the case of GdN. Continuing this theme, much of the modern experimental work has centred around tuning the electronic properties of the RN via nitrogen vacancy concentration, with the conclusions in general being that there is a positive correlation between the nitrogen pressure during preparation and the resulting resistivity of these films~\cite{Holmes-Hewett2020,Shaib2021,Devese2022}. 

To date, the fullest computational treatment of the series as a whole, and the only computational treatment of some of the members, dates back to 2007~\cite{Larson2007}. In recent years there has been a comprehensive computational study~\cite{Petit2010} and following topical review~\cite{Petit2016} covering 140 rare-earth mono-pnictides and -chalcogenides, focusing on the electronic structure and the effects of the strongly correlated 4\textit{f} states. Reference~\cite{Petit2010} predicts most of the series to be small bandgap semiconductors with CeN, PrN, SmN, EuN, TmN and YbN as semi-metals, and EuN as metallic with some heavy Fermion character. Reference~\cite{Petit2016} then discusses the range of computational studies on the RN concluding that these studies in general capture the complexities of the electronic structure of the RN but lack the level of precision needed to determine the true ground state for most of the members, i.e., semi-metallic or semi-conducting. Another recent review~\cite{Wachter2016} focusing on GdN bulk crystals concludes a semi-metallic ground state. 

Computational studies of many of the members are relatively approachable given the simple primitive unit cell of the RN and the availability of pseudo-potentials for the rare-earth elements which were developed using the RN as a test case~\cite{Topsakal2014}. Even so, the usual treatment of the exchange correlation functional (LDA or GGA) results in an underestimation of the bandgap, as is common in semi-conductors, and the strongly correlated 4\textit{f} states require careful treatment. Various methods of treating the 4\textit{f} states have been applied to the RN such as SIC-LDA~\cite{Aerts2004,Petit2010}, the use of hybrid functionals (HSE~\cite{Schlipf2011}, B3LPY~\cite{Doll2008}), semi-empirical quasiparticle correction~\cite{Lambrecht2000}, QSGW~\cite{Chantis2007,Cheiwchanchamnangij2015}, DFT$+U$~\cite{Larson2006,Larson2007,Duan2005,Abdelouahed2007,Holmes-Hewett2025,holmes-hewett2023,Holmes-Hewett2021} and DFMT~\cite{Richter2011}. Quite recently, DFMT was applied to many of the series~\cite{Galler2022} which provides the most recent comprehensive treatment of the electronic structure. An extensive description of the different computational treatments applied to the RN up to 2013 is available in Reference~\cite{Natali2013}, and Reference~\cite{Galler2022} provides details on the developments up until 2022. 

\begin{figure*}
    \centering
    \includegraphics[width=1\linewidth]{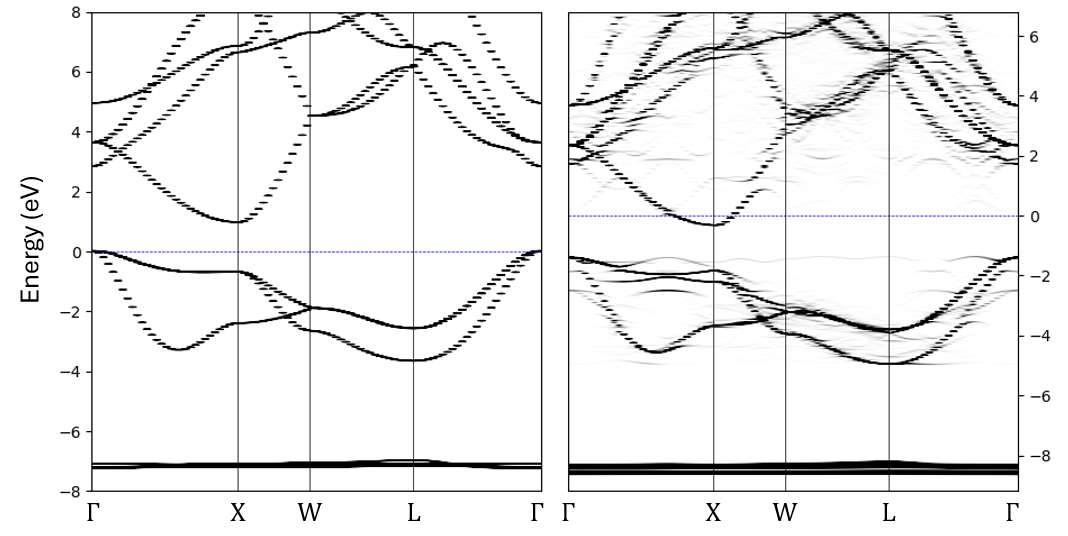}
    \caption{Calculated band structure for stoichiometric LuN (left) and nitrogen vacancy doped LuN (right). The calculations were undertaken in Quantum Espresso v7.2~\cite{QE,QE_2} using the parameters reported in Reference~\cite{Devese2022} and pseudo-potentials from Reference~\cite{Topsakal2014}.}
    \label{bands}
\end{figure*}

\subsection{Electronic Structure of Single Cation RN}

The band structure of the stoichiometric RN in general has valence states formed from N~2\textit{p} bands, which find a maximum at $\Gamma$ where they are degenerate in the cubic rock-salt structure. This three-fold degeneracy is broken along the $\Gamma-$X line, where the $p_z$ state falls sharply in energy and the $p_{x,y}$ states remain doubly degenerate, falling in energy towards X and L. In general, the maximum of the valence band is found at $\Gamma$~\cite{Larson2007}, although some of the lighter members are predicted to have a maximum at X~\cite{Larson2007,Galler2022}. The conduction band is formed from the R~5\textit{d} extended states which have a minimum at X. The largest variation in the electronic structure of different RN relates to the R~4\textit{f} states. Depending on the filling of the given lanthanide, these thread though the conduction and valence bands, hybridising with the R~5\textit{d} and N~2\textit{p} states. It is these 4\textit{f} states which give the individual members their character.

The simplest cases are LaN and LuN with empty and full 4\textit{f} shells respectively. The electronic structure of rock-salt LaN has been the subject of recent computational studies~\cite{Rowberg2021,Deng2021,Yan2014,Banik2023,Sreeparvathy2015}, which concluded that LaN is a small bandgap semiconductor, whose electronic properties can be tuned via nitrogen vacancies, or substitutional oxygen doping. In addition, References~\cite{Sreeparvathy2015} and~\cite{Banik2023} investigate the thermoelectric properties of LaN and suggest that it may be useful for low-temperature thermoelectric applications. The electronic structure of LuN has also been the focus of recent computational studies~\cite{Sreeparvathy2015,oualdine2018,Singh2015,gupta2013,Devese2022}. Reference~\cite{Devese2022} presents a combined experimental and computational study which focuses on identifying the defect states relating to nitrogen vacancies, where it is clear that nitrogen vacancies act as n-type donors. When a nitrogen vacancy is formed in LuN the Lu ions in the material retain their $3+$ charge state. There are then three electrons, which otherwise would have been housed on the nitrogen ion, to be accounted for. Two of these find largely dispersion-less states localized to the vacancy site, the third defect state is higher energy, above the conduction band minimum, thus the third electron lifts the Fermi energy into the Lu~5\textit{d} conduction band. Figure~\ref{bands} shows the band structure for stoichiometric and nitrogen vacancy doped LuN. This gives one mobile electron per nitrogen vacancy which is available for transport. Both optical and electrical transport measurements on LuN thin films have shown the presence of these states with the former showing the development of a low-energy absorption characteristic of free carriers, and the latter shows a resistivity spanning 10 orders of magnitude, changing with the nitrogen pressure during deposition. 

The energy of formation for nitrogen vacancies has been investigated in GdN~\cite{Punya2011,Porat2024}, SmN~\cite{Porat2024} and NdN~\cite{Aravindh2019} (for 2D NdN). These are reported as 3.05~eV~\cite{Punya2011}, 2.60~eV\cite{Porat2024} for GdN, 0.71~eV for SmN and -5.34~eV for NdN, the trend indicates that vacancies will form significantly more easily in the lighter members of the series. The negative value for NdN implies that at least the first nitrogen vacancy will spontaneously form in the 2D structure. Although direct measurement of the nitrogen vacancy content in the RN has so far not been possible, two recent studies have made progress towards a qualitative method to determine the nitrogen vacancy content of deposited thin films. References~\cite{Melendez-Sans2024} and~\cite{Pereira2023} use the relative intensity of the N~1\textit{s} and R~4\textit{d} core level spectra of HoN and SmN measured by X-ray photoelectron spectroscopy to estimate the stoichiometry. They find that films grown at a lower N$_2$ pressure, or at higher temperatures exhibit a lower nitrogen concentration. Using this method they report vacancy concentrations on the order of 10~\% under low pressure or elevated temperature, noting that the spectra only probe the first few monolayers of the deposited films. 

Moving from the full and empty 4\textit{f} shells of LaN and LuN, the next simplest case is GdN, where the half filled 4\textit{f} shell has majority spin states only $\approx$~7.8~eV below the Fermi energy~\cite{Leuenberger2005}. The intra-ion 4\textit{f}-5\textit{d} exchange then results in spin-splitting in the conduction band. The electronic properties of GdN have been studied by several groups. Interestingly, there is some uncertainty regarding the optical gaps, with various reports in the literature. The onset of optical absorption in the ferromagnetic phase represents the majority spin gap, as in GdN both the conduction band minimum and valence band maximum are majority spin. There is general agreement in the existing experimental studies regarding the 0.9~eV majority spin gap~\cite{Trodahl2007,Vilela2024,Azeem2016,Yoshitomi2011}. There is, however, much less agreement regarding the minority spin gap and associated spin splitting (i.e., the difference between majority and minority spin optical gap). Measurements of the spin splitting are reported variously as 0.16~eV~\cite{Vidyasagar2012,Vidyasagar2014}, 0.4~eV~\cite{Vilela2024} and 0.8~eV~\cite{Azeem2016}. In the context of majority-minority spin gaps, the paramagnetic gap can be estimated as an average of the two. The paramagnetic gap is also variously reported from 1.3~eV~\cite{Trodahl2007} to 1.6~eV~\cite{Vilela2024}. 

Nitrogen vacancy dopants in GdN~\cite{Punya2011,Holmes-Hewett2025} act in a similar manner to in LuN, with the difference being the presence of a spin-split conduction band. In GdN, the occupied defect states are one each of majority and minority spin state, while the Fermi energy is lifted into the Gd~5\textit{d} conduction band. Significantly, the exchange splitting of the 5\textit{d} states results in majority spin-only conduction for doping levels up to $\sim$~1$\times$10$^{21}$cm$^{-3}$~\cite{Trodahl2017}. This complete spin polarization of GdN, combined with the orders of magnitude conductivity control via nitrogen vacancy doping~\cite{Lee2015,Holmes-Hewett2020,Devese2023,Shaib2021}, is attractive in a cryogenic spintronic setting. Electron transport in heavily electron doped GdN films has been studied~\cite{Maity2018,Maity2020}, with these, and many other studies, reporting a peak in the magnetoresistance near the Curie temperature, which is understood as a strong scattering of of electrons resulting from magnetic fluctuations in this temperature range, along with changes to the band structure associated with the ferromagnetic transition~\cite{Maity2020,Mi2013}. Very recently a study of films of excellent crystal quality has been conducted reporting a mobility of 130~cm$^{2}$(Vs)$^{-1}$~\cite{Trewick2025}. In addition to nitrogen vacancies there are other possible electron donors, for example Hf, although there are no reports of this being implemented. Acceptor doping has been investigated using Mg~\cite{Lee2015}. These films showed a conductivity which decreased strongly with Mg concentration, understood as a combination of acceptor doping, and an enhanced energy of nitrogen vacancy formation in the Mg-doped films~\cite{Lee2018}.

Although many of the series have seen both computation and experimental interest, it is SmN that stands out in terms of recent interest regarding thin films. While GdN is largely considered in terms of its potential applications, SmN houses complex behaviours relating to strong correlations, inspiring more fundamental interest. The Sm ion in stoichiometric SmN has five electrons in the 4\textit{f} shell, two fewer than the half filled shell of GdN. The five occupied 4\textit{f} states are centred ~$\sim$6~eV below the Fermi energy with a spread of a few eV. The two unfilled majority spin states thread through and hybridize with the Sm~5\textit{d} states which form the conduction band. It is these states, in the presence of nitrogen vacancies, which are suggested to facilitate electron transport in SmN. The conduction band of SmN, like GdN, is spin polarized via the intra-ion 4\textit{f}-5\textit{d} exchange. Optical absorption measurements, so far only conducted in the paramagnetic phase, show an absorption at 1.2~eV relating to the N~2\textit{p}$\rightarrow$Sm~5\textit{d} transition~\cite{Azeem2018,Holmes-Hewett2019}, similar to that of DyN~\cite{Holmes-Hewett2020,Azeem2013a}, and GdN~\cite{Trodahl2007,Vilela2024,Azeem2016,Yoshitomi2011,Vidyasagar2012,Vidyasagar2014,Holmes-Hewett2025}

As discussed above, the presence of nitrogen vacancies in LuN and GdN results in localized defect states which thread though the intrinsic band gap region, and the Fermi energy is lifted into the extended state 5\textit{d} conduction band, accommodating one mobile electron per nitrogen vacancy. In SmN the situation appears quite different. Although similar defect states appear in the intrinsic gap region in SmN, the electrons in these states are not localized to the vacancy site, but rather shared amongst the majority spin 4\textit{f} states on Sm ions neighbouring the vacancy~\cite{Holmes-Hewett2021}. As there are two empty majority spin states per Sm ion, and six neighbours to each vacancy, these states pin the Fermi energy to the intrinsic gap region. The defect states and the pinned Fermi energy result in an optical absorption spectrum quite distinct from GdN and LuN, with a strong absorption in the intrinsic gap region which is enhanced with increased nitrogen vacancy doping~\cite{Holmes-Hewett2019,holmes-hewett2023}. The contrasting nature of the electronic transport in SmN, compared to GdN and LuN is also evident in an enhanced anomalous Hall effect. Measurements of the anomalous Hall effecnt in SmN~\cite{Holmes-Hewett2018}, show an enhanced magnitude compared to GdN~\cite{Trodahl2017} consistent with a 4\textit{f} mediated transport channel. As discussed above, Reference~\cite{Melendez-Sans2024} has examined the nitrogen content of SmN thin films though XPS. They compare the XPS Sm 3\textit{d} core level spectra of SmN to Sm metal. The latter has a divalent surface so clearly shows features relating to both the Sm$^{3+}$ 3\textit{d}$^9$4\textit{f}$^5$ and Sm$^{2+}$ 3\textit{d}$^9$4\textit{f}$^6$ multiplets. They find the the Sm$^{3+}$ 3\textit{d}$^9$4\textit{f}$^5$ configuration is dominant in all SmN films, with a weak indication of the Sm$^{2+}$ 3\textit{d}$^9$4\textit{f}$^6$ configuration appearing in the most nitrogen-deficient films. 

\begin{figure}
\centering
\includegraphics[width=\linewidth]{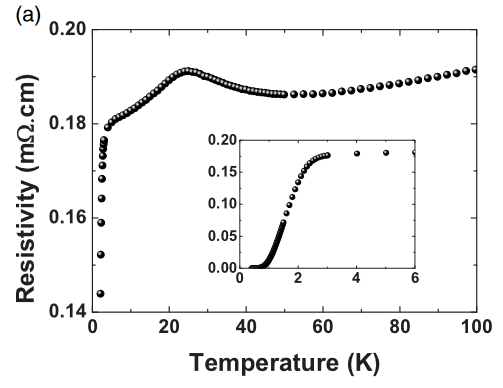}
\caption{Temperature-dependent resistivity of a SmN film, showing an anomaly near the Curie temperature of 27 K and the onset of superconductivity below about 3 K. The inset shows the full superconducting transition. The Figure has been reproduced with permission from Reference~\cite{Anton2016}.}
\label{SC}
\end{figure}

Significantly, SmN has also been observed to harbour a superconducting state~\cite{Anton2016} when appropriately doped with nitrogen vacancies~\cite{holmes-hewett2023}. The superconducting transition temperature of $\sim$2~K is well below the ferromagnetic transition temperature of $\sim$27~K implying these states may coexist. Figure~\ref{SC} shows a resistivity measurement conducted on a SmN thin film displaying both the ferromagnetic transition (anomaly near 30~K) and the superconducting transition near 2~K. Further to this, an enhanced superconducting transition temperature has been observed in GdN/SmN multi-layers, the nature of the enhancement, or the origin of the superconducting state itself, is not yet understood. References~\cite{holmes-hewett2023} and~\cite{Anton2016} argue that the low transition temperature, coexistence with ferromagnetic order (yet near-zero internal magnetic field), and the apparent spin polarization of the defect states imply an unconventional nature of the superconductivity. 

Of the other members studied, ErN has seen some recent attention regarding its electronic structure~\cite{Upadhya2021,McKay2020,Upadhya2022}. Er-based compounds are of specific interest due to the intra-4\textit{f} transitions near 1.54~$\mu$m which are in the telecommunications range, thus Er-doped III-nitride materials are studied for applications relating to this purpose~\cite{Sun2016,Steckl2002,Dahal2009}. Two recent studies of ErN detail the band structure of the material in the context of the Er:III-nitride systems. Reference~\cite{McKay2020} reports a photo-luminescence study of ErN thin films yielding a direct band gap of 0.98~eV. This was followed by a resonant-photoemission study~\cite{Upadhya2022} detailing the N~2\textit{p} valence band maximum and locating the Er~4\textit{f} states at $6-8$~eV below the Fermi energy. These results are corroborated by a DFT based computational study in the same report. 

\subsection{Electronic Structure of Mixed Cation (R,R')N}

\begin{figure}
\centering
\includegraphics[width=\linewidth]{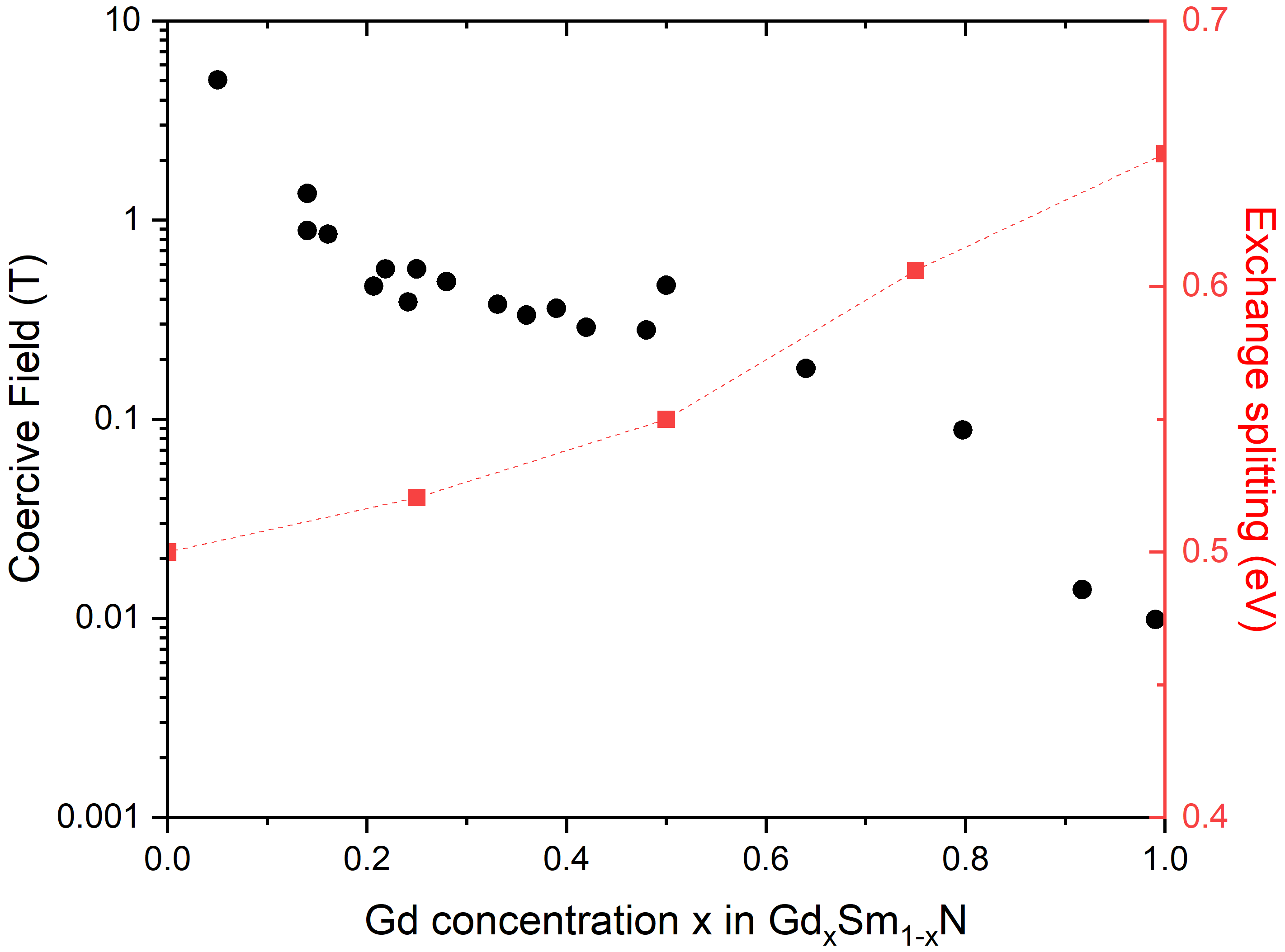}
\caption{Experimental coercive fields measured at 5~K (black circles), and calculated exchange splitting between the majority and minority spin bands (red squares), in various Gd$_x$Sm$_{1−x}$N compositions. The figure has been reproduced with permission from Reference~\cite{Porat2024}.}
\label{Ex-cf}
\end{figure}

As with the magnetism there is again significant potential for application regarding the electronic properties of alloys of the RN. Again (Gd,Sm)N is one of very few combinations which has seen any attention. As discussed above in section~\ref{S:Magnetic:Alloys} the substitution of Sm ions into GdN does not interrupt the magnetic ordering, but does have a significant effect on the coercivity of the material. 

A recent study of the optical properties and electronic structure of (Gd,Sm)N~\cite{Porat2024} shows that the extended state 5\textit{d} bands on the lanthanide cations in (Gd,Sm)N fully hybridise while the localized 4\textit{f} do not for the most part. The result is an optical band gap which interpolates between those of GdN and SmN with cation concentration. The spin-splitting in the conduction band of these materials is a result of the intra-ion exchange between the localized 4\textit{f} states and the 5\textit{d} states. This exchange is a result of the spin-polarization of the 4\textit{f} states and thus is expected to be similar in GdN and SmN, which is significant considering their vastly contrasting magnetic properties~\cite{Meyer2008,Ludbrook2009}. The computational study shows the spin-splitting, like the optical band gap, interpolates between the end members, such that the exchange splitting is relatively constant while the coercivity can be tuned (see Figure~\ref{Ex-cf}). This has potential application in devices which are sensitive to the exchange splitting, rather than the magnetic moment of a material, such as the magnetic Josephson junctions used in superconducting spintronic devices~\cite{Birge2024}. The only other (R,R')N alloy to be studied experimentally is (Gd,Dy)N~\cite{Pot2021} with only a brief mention of the electronic properties, which in general resemble those of the end members, GdN and DyN. A computational study of the structural, electronic, magnetic and elastic properties of (La,Gd)N alloys~\cite{Mahfoud2018} has been undertaken using the full potential linear muffin-tin orbital method with a Hubbard \textit{U}$_f$ correction to the 4\textit{f} electrons. Regarding the electronic structure, the study finds all concentrations of (La,Gd)N to be metallic, although this report states that the lack of a bandgap is likely due to the underestimate common to LDA methods, and in this case it is untreated with a \textit{U}$_d$ parameter as discussed above. 

\section{Applications in Cryogenic Electronics}

\label{Applications}

To date, various structures leveraging the tunable electronic and magnetic properties of the RN have been demonstrated for application in cryogenic electronics. In the field of superconducting electronics one of the key challenges is the need for devices which will enhance the density of existing memory units, while remaining operational at 4~K and maintaining compatibility with the present superconducting technologies i.e. single flux quantum logic (SFQ), and its derivatives~\cite{Soloviev2017,Alam2023,IRDS2023,Mukhanov2011}. The recent interest in the development of scalable cryogenic quantum computing has provided another driver for such a memory system, which could also be exploited in support of the superconducting control processor located on a cryogenic stage adjacent to the qubits. An SFQ-based unit has been proposed to facilitate the scale up of quantum computing systems, i.e. to provide alternative digital control of qubit states in place of more standard analogue microwave-based solutions.

In the present section we discuss the application of the RN materials in various devices, mostly regarding cryogenic memory. We begin by discussing resistive memory devices, then the integration of RN materials with superconductors, again focusing on memory devices. Finally, we discuss the use of the RN materials as the barrier layer in Josephson junctions, here we discuss some of the underlying physics and applications in logic elements for quantum computing. 

\subsection{Resistive Memory Devices}

The tunable electronic and magnetic properties of the RN have already been explored in various memory device architectures. The first exploration of such a device was in the form of a magnetic tunnel junction (MTJ). In these structures two ferromagnetic electrodes are separated by a non-magnetic, insulating, tunnel barrier. Memories based on these MTJs in general have great potential to outperform charge-based memories. They are non-volatile, dense, have low switching time and high endurance. Although not necessarily required they also have a long retention time on the order of years~\cite{Bhatti2017}. The memory state of the device is stored in the relative orientation of the fixed (high coercivity) and free (low coercivity) ferromagnetic layers. The readout is via an electrical measurement through the tunnel barrier. In the parallel configuration both layers have a high density of majority spin states at the Fermi energy, while in the anti-parallel configuration there is a reduced population of majority spin states in one layer. This contrasting density of states results in a higher tunnelling resistance for the anti-parallel configuration. The RN materials are attractive for forming the ferromagnetic layers in such a device as the strong spin polarization should in principle yield a very large resistance contrast in the two configurations. 

\begin{figure}
    \centering
    \includegraphics[width=1\linewidth]{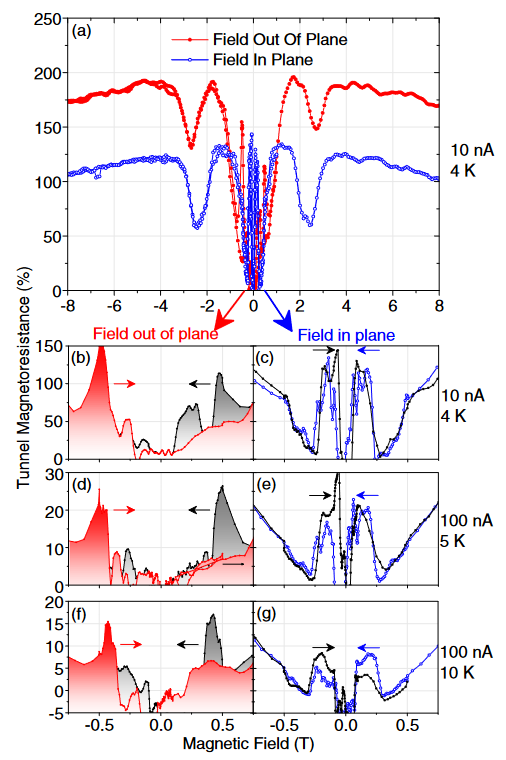}
    \caption{(a) Magnetoresistance data from a GdN/AlN/SmN tunnel junction obtained from −8 T to 8 T at 4 K in both the in-plane and out-of-plane orientations with a measurement current of 10 nA. (b)–(g) Tunnel magnetoresistance data at low fields measured both in increasing and decreasing fields using different currents and temperatures. The figure is reproduced with permission from Reference~\cite{Warring2016a}.}
    \label{TMR}
\end{figure}

MTJ structures comprising combinations of GdN, SmN and DyN ferromagnetic layers with GaN, AlN and LuN barrier layers have been investigated~\cite{Devese2023,Miller2020,Warring2016a}. Here we highlight GdN/LuN/DyN~\cite{Devese2023} and GdN/AlN/SmN~\cite{Warring2016a} for the contrast of spin/spin and spin/orbital dominated magnetic materials. GdN/LuN/DyN features two spin-dominated magnetic materials, so fits the conventional description of MTJs above. Here LuN is used as the non-magnetic tunnel barrier. The use of LuN enables the growth of fully lattice matched hetero-structures, and if combined with the recent advancements in thin film deposition described in Section~\ref{growth} could result in well-ordered, fully-epitaxial MTJs. In the case of Reference~\cite{Devese2023} epitaxial growth was not used. These devices, made entirely of insulating layers, only showed a very modest $\sim1$\% contrast of resistance at zero field between the two magnetic configuration states, but did show a development of non-linear current-voltage characteristics indicating that the insulating LuN layer was performing as an effective tunnel barrier.

\begin{figure}
\centering
\includegraphics[width=\linewidth]{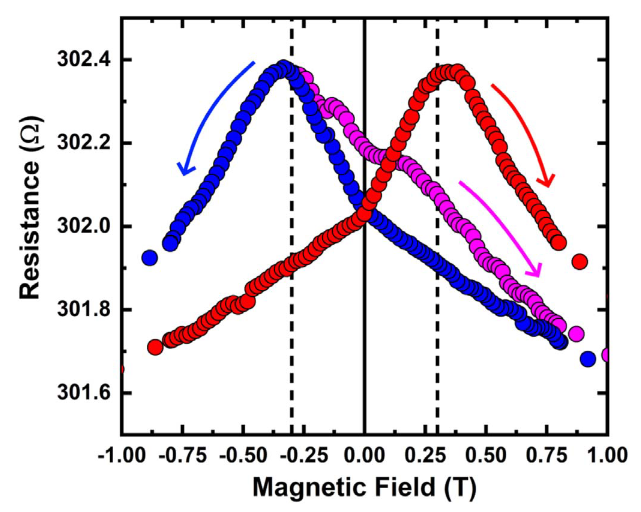}
\caption{Measured CIP magneto-resistance in a GdN/Lu/DyN device, The red and blue data show peaks corresponding to the coercive field of the free layer of the device. The pink data demonstrates a working loop for the device. The figure has been reproduced from Reference~\cite{Devese2023}.}
\label{GMR}
\end{figure}

The orbital-dominated magnetism of SmN in the GdN/AlN/SmN~\cite{Warring2016a} magnetic tunnel junctions makes the behaviours more complex. Here, when the two ferromagnetic layers are aligned with an applied field, the spin magnetic moments are opposed. This creates the opposite effect to that of conventional MTJs, so the high resistance state is now found in the \textit{parallel} magnetic configuration, and the low resistance state is found in the \textit{anti-parallel} configuration, when the spin-moments of the materials are aligned. The resistance contrast between the two configurations in Reference~\cite{Warring2016a}, shown in Figure~\ref{TMR}, was roughly 200~\%, significantly higher than in Reference~\cite{Devese2023}. 

With ferromagnetic layers made of semiconducting RN rather than the metallic transition-metal-based layers in conventional MTJs, achieving ohmic electrical contacts to RN structures becomes less trivial and was investigated with several metallic electrodes~\cite{Miller2020,Ullstad2015}. Reference~\cite{Miller2020} found that Ohmic contact was facilitated by Au and Al electrodes while Gd electrodes resulted in a Schottky barrier forming at the Gd/GdN interface. An earlier report~\cite{Ullstad2015} which investigated vertical transport though Au/GdN and Ag/GdN structures also found Ohmic contact was formed with these metallic electrodes. 

RN devices based on the giant magnetoresistance (GMR) effect in the current-in-plane (CIP) geometry have also been investigated. In CIP-style devices current is passed along a very thin layer of non-magnetic conductor, which itself is bounded by two ferromagnetic layers. Again the information is stored in the relative alignment of the two ferromagnetic layers, and readout is via a current passed along the central conductive layer. An enhanced resistance is present in the anti-parallel configuration, a result of the enhanced spin scattering resulting from this configuration. Conventional GMR-CIP-style devices use metallic materials for both the ferromagnetic and non-magnetic layers. The largest contrast is found when current is restricted to the non-magnetic layer, so to promote this the thickness of the ferromagnetic layers are reduced. The use of ferromagnetic insulators simplifies this as there is an intrinsic resistivity contrast, rather than an engineered resistance contrast, between the non-magnetic conductor and the ferromagnetic layers. To date, there is only one demonstration of such a device which comprises GdN/Lu/DyN layers~\cite{Devese2023}. The GMR-CIP device in Reference~\cite{Devese2023} displays a clear non-volatile memory, although again there is only a modest contrast between the two configurations at zero field, shown in Figure~\ref{GMR}. 

\subsection{Integration with Superconductors}

\begin{figure}
    \centering
    \includegraphics[width=1\linewidth]{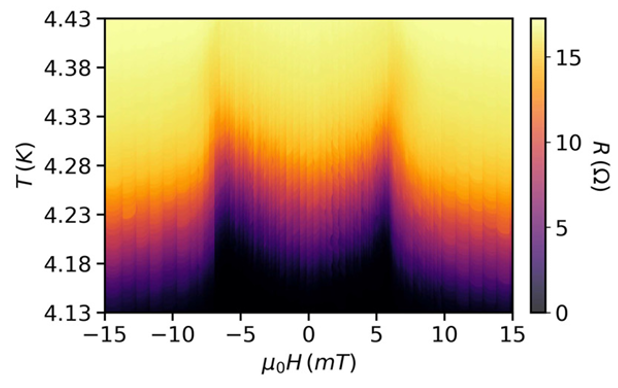}
    \caption{Resistance of Nb (8~nm)/GdN (3~nm) versus the applied field and temperature showing control of the superconducting transition. The figure is reproduced with permission from Reference~\cite{Banerjee2023}.}
    \label{Dom-wall}
\end{figure}

Conventional superconducting memory technologies are formed from non-magnetic Josephson junctions, where the working principle is relatively straightforward. Superconducting loops store information in the form of a single flux quantum pulse, which can be detected for readout at a later time. These memories are effective, fast, and efficient, but scalability is a serious issue as the physical requirements result in memory elements on the scale of at least 10$\times$10~$\mu$m$^2$~\cite{Soloviev2017}. The integration of ferromagnetic materials in superconducting circuits can enable various memory devices which can achieve much higher densities~\cite{Alam2023}.

There has been significant work regarding the integration of ferromagnetic materials with superconducting materials. This is due to fundamental and applied interests, motivated by the rich physics in such systems~\cite{Buzdin2005,Bergeret2005}, and the promise of devices with enhanced efficiency and entirely new computation methods. Interest so far has been dominated by systems comprising superconductors and metallic ferromagnets. However, the use of ferromagnetic insulators in such systems, originally considered decades ago~\cite{De-Gennes1966}, has advantages relating to the well-defined interfacial decay of the superconducting wave-function in the ferromagnetic insulator, which will only penetrate to the order of the interatomic spacing. Contrasting the metallic systems, with insulating ferromagnets in a current-in-plane geometry, the electrons forming a superconducting Cooper pair are reflected at the interface, resulting in an effective exchange field in the superconductor~\cite{Tokuyasu1988}. In such a system, when the exchange field in the ferromagnetic insulator is large, this effective exchange can be used to affect the properties of the superconductor. For example, a simple magnetic switch can be developed from bi/tri-layers, which can display an infinite magnetoresistance. In fact, such a device has already been demonstrated using a EuS/Al/EuS system~\cite{Li2013} over a decade ago. 

Bi/tri-layers comprising RN and superconductors, primarily NbN:GdN, have been studied in the context of these interface effects. One of the very first reports on thin films of GdN explored proximity effects in NbN/GdN bilayers, revealing an enhancement of the superconducting T$_C$ with increasing GdN thickness due to confinement of the Cooper pairs in the superconducting NbN, despite the strong ferromagnetism of the GdN~\cite{Xiao1996}. Strong interfacial exchange fields ($\sim1.7~$T) have been observed in NbN/GdN/TiN tri-layers, by means of tunnel junction spectroscopy of the NbN layer~\cite{Pal2015}. For the case of GdN this interfacial exchange, on the order of 1~T, can be moderated by the application of a small magnetic field of about 100~Oe. This is demonstrated in Reference~\cite{Banerjee2023} where the transition temperature of a thin Nb layer is affected to the order of $\sim$100~mK through the application of an external field of only $\sim10~$mT. The control of transition temperature is shown in Figure~\ref{Dom-wall}. In both Reference~\cite{Pal2015} and~\cite{Banerjee2023}, the origin of the effect is seen to be the paramagnetic pair-breaking at the GdN:Nb interface, rather then the orbital pair-breaking due to the stray field of the GdN. 

\begin{figure}
    \centering
    \includegraphics[width=1\linewidth]{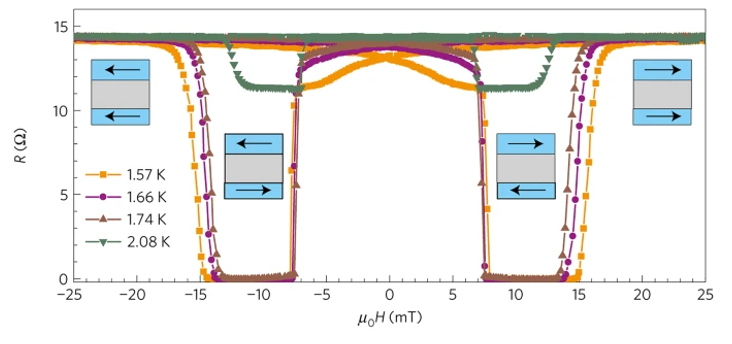}
    \caption{Resistance as a function of applied magnetic field for GdN/Nb/GdN pseudo-spin valves. The Figure is reproduced with permission from Reference~\cite{Zhu2016}.}
    \label{Exchange-coupling}
\end{figure}

Another study~\cite{Zhu2016} reports tri-layers of GdN/Nb/GdN where there exists a superconducting exchange-coupling between the GdN layers, mediated by the superconducting state of the Nb layer. Similar to the EuS/Al/EuS device discussed above, this GdN/Nb/GdN device shows an effectively infinite magnetoresistance on application of $\sim$10~mT, showing great promise for a memory device. The clear switching between the resistive and non-resistive states can be seen in Figure~\ref{Exchange-coupling}. A recent preprint~\cite{Dutta2023} extends this concept further and demonstrates the reciprocal effect of the magnetic domains and the superconducting state in GdN/Nb/GdN superconducting spin-valves. The authors suggest the potential for using the multiple reproducible resistive states achieved with the field-history-dependent magnetic domain structures in cryogenic neuromorphic computing.

Reference~\cite{Cascales2019} reports a superconducting spin-valve type device where the memory element is formed from GdN/Nb/GdN tri-layers. In this configuration the thickness of the Nb layer is reduced below the superconducting coherence length (in this case the Nb is 10~nm thick), which results in the strong interfacial exchange interaction from the adjacent GdN layers combining in the superconducting Nb. In this geometry, the anti-parallel magnetic configuration is the ground state of the device, with a critical current on the order of~120~$\mu$A at 1.3~K. When the device is switched into the parallel configuration, requiring application of a magnetic field of 250~Oe, the critical current shows significant contrast, dropping to $\sim$~60~$\mu$A. In this way, the memory state of the device can be probed via a current pulse applied to the structure with the presence or absence of a voltage indicating the state of the device.

\begin{figure}
    \centering
    \includegraphics[width=1\linewidth]{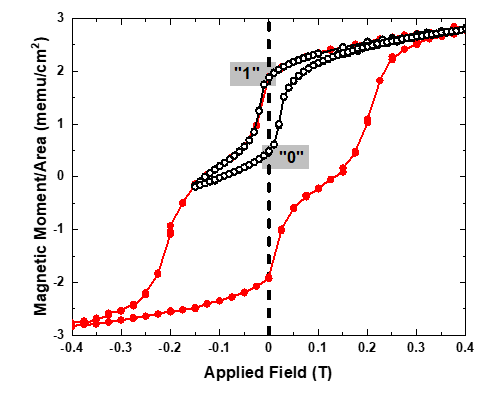}
    \caption{Magnetic measurement on (Gd,Sm)N/Al/(Gd,Sm)N tri-layer structure. The red data shows a full hysteresis loop and the black data shows the working loop where only the free layer is switched. The Figure has been adapted with permission from Reference~\cite{pot2023}.}
    \label{Minor-loop}
\end{figure}

Another device concept forms a pseudo spin-valve from two RN layers, separated by a non-magnetic metallic blocking layer~\cite{pot2023}. The memory element here is similar to conventional tri-layer magnetic memory with the two ferromagnetic layers formed from (Gd,Sm)N. The compositions of these layers are chosen such that they have significantly differing coercivities forming a hard-soft magnetic pair. In addition, the \textit{net} magnetic moment of the two ferromagnetic layers are balanced such that in the parallel configuration there exists a significant fringe magnetic field, while in the anti-parallel configuration the net magnetic moment, and hence the fringe field, is near zero. This forms a micron-scale switchable magnet, the state of which can be detected with an adjacent Josephson junction that is sensitive to the fringe field~\cite{Held2006}. The advantage of the device presented in Reference~\cite{pot2023} is that the proposed electrical readout of the state of the memory element is via a conventional Josephson junction, enabling the direct integration with conventional superconducting electronics. Reference~\cite{pot2023} demonstrates the memory function of the device alone, with the read of the device via a magnetometer and the write accomplished via an external magnetic field, shown in Figure~\ref{Minor-loop}. However, in principle both the read and write operations could be achieved electrically, with the switching of the magnetic configuration done with the field originating from a nearby write line, or by using a spin-transfer or spin-orbit torque applied directly to the switchable magnet structure~\cite{Avci2017,Rowlands2019}.

\subsection{Integration with other Josephson-based Applications}

Finally, a significant body of work has formed around the integration of the RN as the barrier layer in Josephson systems, motivated by a combination of technological and fundamental interest. First, we briefly summarise the relevant physics of Josephson junctions and the effect of incorporating a ferromagnetic barrier to the behaviours of the junctions. 

\subsubsection{Josephson Junction Operation and Magnetic Josephson Junctions made using Ferromagnetic Metals}

The conventional Josephson effect uses BCS (Bardeen-Cooper-Schrieffer) superconducting electrodes separated by a non-superconducting barrier of a few~nm~\cite{tinkham2004,Tafuri2019}. Each superconducting electrode can be described by a condensate of singlet Cooper pair superconducting wavefunctions $\Psi=|\Psi|e^{i\theta}$, where $|\Psi|$ is the wavefunction density and $\theta$ is the phase. Such wavefunctions, overlapping in the barrier, give rise to a phase-difference across the device $\varphi=|\theta_1-\theta_2|$. The two Josephson equations describe the electrical properties:

\begin{equation}
    I(\varphi)=I_c\sin(\varphi)
\end{equation}
\begin{equation}
    d\varphi/dt=\hbar/(2e)V(t) ,
\end{equation}

\noindent where $I_c$ is the maximum supercurrent that the Josephson junction can sustain, and $V(t)$ is the time-dependent voltage drop across the device. 

If an external magnetic field is applied perpendicular to the super-current $I_C$ is modulated as a function of the induced magnetic field as a Fraunhofer, or Airy, pattern, respectively for a square or circular geometry~\cite{tinkham2004}. The periodicity of the critical current as a function of the applied field, $I_c(B)$, finally depends on the effective magnetic area of the device. Specifically, the total magnetic flux in the Josephson junction is $\Phi=\mu_0B A_m$, with $A_m$ the area. The critical current as a function of field $I_c(B)$ is then 

$$I_c(B)=\sin(\pi\Phi/\phi_0)/(\pi\Phi/\phi_0),$$ 

\noindent with $\phi_0$ the magnetic flux quantum. The maximum of the Fraunhofer-like modulation of a non-magnetic Josephson junction occurs at zero magnetic field. When deploying a ferromagnetic material as the barrier in a Josephson junction the presence of a hysteretic magnetization of the barrier must also be taken into account in the functional form of $B$ as 

$$B(H)=\mu_0H+\mu_0M(H).$$

\noindent Therefore, depending on the direction of an applied external magnetic field, the barrier acquires a finite residual magnetization. In order to compensate for this residual magnetization, the Fraunhofer-like $I_c(H)$ curve shows a hysteresis and shifts along the positive field direction when ramping the external magnetic field from negative to positive (i.e., in presence of a negative residual magnetization). The contrary holds in the presence of a positive residual magnetization. Such a hysteretic mechanism has been proposed for the implementation of cryogenic random access memories~\cite{Soloviev2017,Alam2023}. For these applications, several ferromagnetic materials have been explored, mostly relying on itinerant and percolative metallic ferromagnets, like Ni, Py, PdFe~\cite{Madden2019,Birge2018,Ryazanov2012}. 

Current cryogenic electronics employing Josephson devices largely rely on Nb/AlO$_x$-based structures~\cite{Alam2023}. Regarding logic systems, one of the motivations for using AlO$_x$ is the high resistivity this material offers as a barrier layer. The switching time of Josephson junctions in these systems

$$\tau=\frac{\phi_0}{2\pi I_C R_N},$$

\noindent is inversely proportional to the product of $I_C$ and the normal state resistance $(R_N)$. Josephson junctions formed from Nb/AlO$_x$ can have a characteristic voltage $V_C=I_CR_N \approx~1$~mV resulting in a switching frequency on the order THz~\cite{Chen1999}. Recently, an SFQ system based on Nb/AlO$_x$ components has demonstrated simple programmes executed at a frequency of 50~GHz~\cite{Tanaka2023}. However, the intrinsic metallic nature of most ferromagnetic materials provides a much larger barrier transparency, dramatically reducing $R_N$ and providing a low characteristic voltage $V_C~\approx~100~$nV. The switching frequency corresponding to this lower characteristic voltage is orders of magnitude smaller than that of oxide-based devices, on the order 100~MHz. These contrasting time scales make the use of metallic ferromagnetic Josephson junctions largely unsuitable as switchable circuit elements in conventional Nb/AlO$_x$-based Josephson junction logic systems. 

Progress has been made combining the hysteretic nature of ferromagnetic materials with the enhanced normal-state resistance of insulators by forming multilayer Josephson junctions comprising both of these attributes. Of note in this area are multi-layer structures such as Nb/AlO$_x$/Nb/PdFe~\cite{Vernik2013,Larkin2012} and Nb/AlO$_x$/(Nb)/Py/Nb~\cite{Parlato2020}. These devices have seen enhanced switching characteristics compared to conventional metallic Josephson junctions. However, the more involved growth processes and additional interfaces introduce difficulties. The incorporation of insulating ferromagnetic materials to push up the characteristic voltage while maintaining the hysteretic behaviour is less developed, but may enhance the capabilities of ferromagnetic Josephson devices for cryogenic electronics applications.

\subsubsection{Magnetic Josephson Junctions with RN Materials}

The opportunity to employ ferromagnetic insulators as the barrier layer in Josephson junctions is attractive as, if successful, these would provide simple structures intrinsically compatible with conventional Josephson junctions, thus appropriate for application in conventional superconducting logic systems~\cite{Ustinov2003}. In this context, the insulating RN provide an attractive area of study for high characteristic voltage magnetic Josephson junctions, while the possibility of deploying the highly tunable magnetic properties of RN-based alloys provides a fertile playground for both device developments and fundamental studies. 

\begin{figure}
\centering
\includegraphics[width=\linewidth]{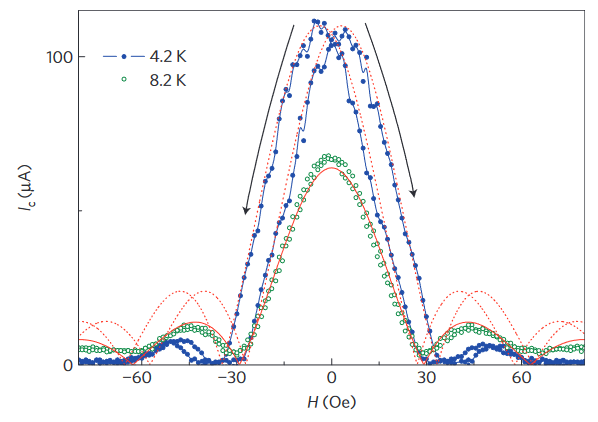}
\caption{In-plane magnetic field dependence of junction critical current for a 5$\pm$2~nm GdN barrier in NbN/GdN/NbN, showing Fraunhofer-like oscillations at 4.2 and 8.2~K. The solid line shows a fit to the 8.2~K data, where the effective field seen by the junction has an additional 10~Oe originating from the barrier moment, which reverses smoothly at low field. The calculated Fraunhofer patterns at 4.2~K (using a constant field-shift) are shown as dotted lines, to emphasize the suppression of side-lobe critical current I$_{c_2}$. The Figure is reproduce with permission from Reference~\cite{Senapati2011}.}
\label{frau}
\end{figure}

The first exploration of any of the RN in Josephson junction structures comprised NbN/GdN/NbN devices~\cite{Senapati2011}. These results displayed a tunnelling current and a spin-filter effect resulting from the exchange-split 5\textit{d} conduction band of GdN, while a hysteretic Fraunhofer pattern (Figure~\ref{frau}) demonstrated the effect of the ferromagnetism, resulting from the 4\textit{f} electrons and their influence on the transport properties through 4\textit{f}-5\textit{d} exchange. Significantly, the characteristic voltage $(V_C=0.3$~mV) was enhanced compared to that in metallic ferromagnetic Josephson junctions, showing clear promise for the integration of these structures with conventional Josephson-junction-based devices. Additional studies on similar structures have supported the tunnel nature of NbN/GdN/NbN Josephson junctions through evidence of double-Schottky barriers at the NbN/GdN interfaces, where the resulting depletion regions in the GdN affect the magnetic and spin-filter properties of the structures~\cite{Pal2013}. 

\begin{figure}
    \centering
    \includegraphics[width=\linewidth]{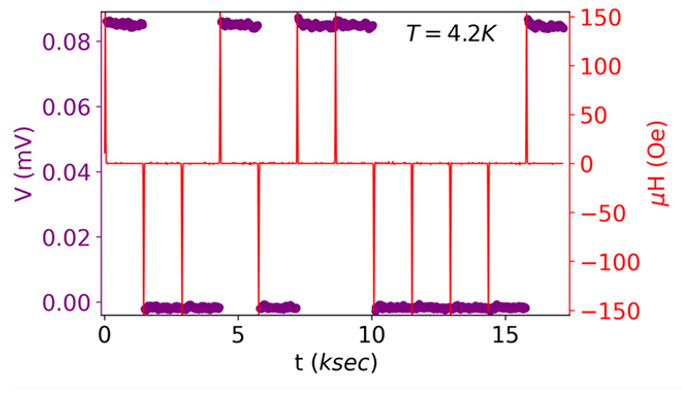}
    \caption{Switching between zero and finite voltage states at zero field by application of positive and negative magnetic fields in a NbN/GdN/NbN stack. The Figure is reproduced with permission from Reference~\cite{Sharma2024}.\color{black}}
    \label{SC-switch}
\end{figure}

Recently there has been a demonstration of a memory element comprising a NbN/GdN/NbN Josephson junction structure. As discussed above, the enhanced I$_c$R$_N$ product in the insulating ferromagnetic barrier layer results in a switching time comparable to that of conventional Josephson Junction based logic devices, allowing direct integration. The device reported in Reference~\cite{Sharma2024} features a single layer GdN long-Josephson junction. Although a single-layer junction will display a hysteresis, manifesting as a displacement of the current voltage characteristics in magnetic field, both magnetic orientations of the barrier layer still yield the same resistance value at zero field, thus it is not useful as a memory device. Here this issue has been addressed by extending the junction dimension (L) perpendicular to the in-plane magnetic field direction such that L/$\lambda_J$ is $>$ 1, $\lambda_J$ being the Josephson penetration depth. When L/$\lambda_J$ is $>$ 1 the self-field generated by the current in the junction is no longer negligible and thus the effective magnetic flux ($\phi_{eff}$) at zero field is formed from

$$\phi_{eff}=\phi_{self} \pm \phi_{M},$$

\noindent where $\phi_{self}$ and $\phi_{M}$ are the magnetic flux from the self-field and barrier magnetization respectively. The requirement for a contrasting memory state at zero field is found here as the sign of $\phi_{M}$ depends on the orientation of the magnetization of the barrier layer, while  $\phi_{self}$ does not. In this regard, Reference~\cite{Sharma2024} shows a functional memory device formed from a single Josephson junction of area $20 \times 20~\mu$m, and with a I$_c$R$_n$ product of $\sim$~0.2~mV. Figure~\ref{SC-switch} shows the device switching between the superconducting and resistive states upon application of a $\pm 150$~Oe applied field. 

An additional application of the NbN/GdN/NbN long-Josephson junction is as a zero-field and dissipation-less superconducting diode. Diode operation has been demonstrated at speeds of 28~GHz down to temperatures of 4.2~K with an efficiency tunable by setting the magnetic microstructure of the GdN~\cite{Sharma2023}.

\begin{figure}
\centering
\includegraphics[width=\linewidth]{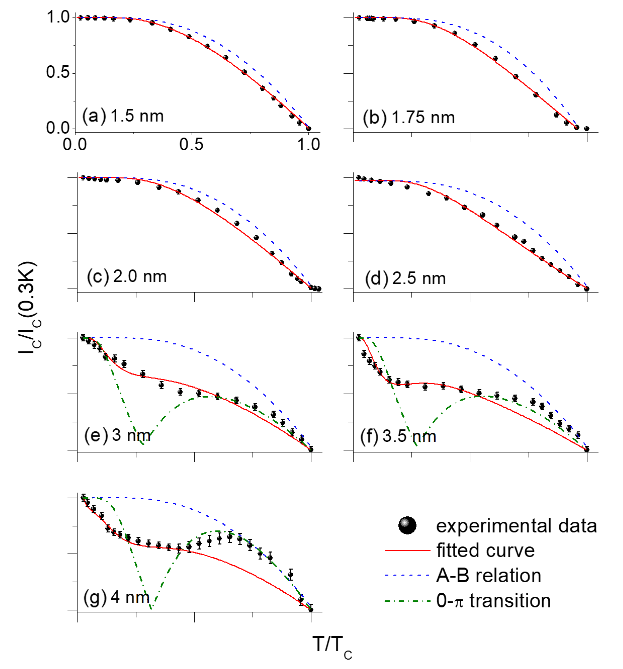}
\caption{Black dots: Measured critical current as a function of temperature for various NbN/GdN/NbN based Josephson junctions. Blue dashed lines: Ambegaokar-Baratoff relation plotted for comparison. Red lines: fitted curves. The figure has been reproduced with permission from Reference~\cite{Caruso2019}.}
\label{0-pi}
\end{figure}

Not only do NbN/GdN/NbN Josephson junctions allow, in principle, magnetic memories and other devices for cryogenic electronics, but their study also provides a rich platform to understand unconventional Josephson mechanisms, possibly allowing for novel applications. In Reference~\cite{Massarotti2015}, the first evidence of macroscopic quantum tunnelling in a hybrid ferromagnetic Josephson junction has been reported. These results are particularly relevant in superconducting electronics applications where a substantial under-damping is required (e.g. superconducting quantum technologies). These results add to those discussed earlier in Reference~\cite{Pal2014}, where the first evidence of higher-order harmonics in the current-phase relation has been reported down to 4~K. Further characterization of NbN/GdN/NbN Josephson junctions with different barrier thicknesses down to 300 mK in Reference~\cite{Caruso2019} has demonstrated that an unconventional 0-$\pi$ transition occurs, observed here in the critical current as a function of temperature. Indeed, rather than the typical cusp corresponding to the 0-$\pi$ transition measured in standard metallic SFS Josephson junctions~\cite{Buzdin2005}, a plateau or a non-zero local minimum in the $I_C(T)$ curves has been measured for certain GdN thicknesses, shown in Figure~\ref{0-pi}.

Here, the presence of the unconventional 0-$\pi$ transition is modelled by considering the transport to comprise two separate channels, which relate to the multi-domain structure in the barrier $($area~$7\times7~\mu$m$)$. This qualitatively mimics the occurrence of spin-mixing effects, and has been further developed in Reference~\cite{Ahmad2022}. Indeed, the combined presence of spin-mixing and spin-filtering effects has been widely discussed to provide generation of unconventional spin-triplet superconductivity~\cite{Eschrig2011}. 

In Reference~\cite{Ahmad2022}, a microscopic modelling of these devices has been proposed to demonstrate the correlation between the unconventional thermal behaviour of NbN/GdN/NbN Josephson junctions and the arising of spin-triplet superconductivity. First, it has been demonstrated that the unconventional 0-$\pi$ transition in Ref.~\cite{Caruso2019} is more likely described as a broadened 0-$\pi$ transition in temperature, extended over a range of few~K, thus allowing the possibility to operate these junctions as controllable $\varphi-$Josephson junctions in temperature. Secondly, the NbN/GdN/NbN Josephson junctions complex current-phase relation is strongly related to the presence of spin-triplet superconducting processes, governed by the magnetic behaviours of the barrier. Specifically spin-orbit coupling, the magnetic inhomogeneities, and impurities in the barrier play a fundamental role. Finally, it has been demonstrated that an external magnetic field allows the control of the relative contribution of each of these effects, thus providing an experimental handle to tune spin-triplet and spin-singlet transport, behaviour which may be applied in spintronic devices. 

Other than GdN, only DyN has been studied in a Josephson junction geometry~\cite{Muduli2014}. In this report, concerning NbN/DyN/NbN structures, coherent tunnelling of Cooper pairs is found at a barrier thickness of $\sim$~1~nm, while above this thickness there was evidence of spin filtering effects and conventional tunnelling behaviours. 

\subsubsection{Potential for RNs in Superconducting and Quantum Logic}

The use of insulating ferromagnetic materials in Josephson junctions is becoming of greater interest as developing technologies push conventional materials to their limits. In addition to integration with conventional superconducting logic devices, high characteristic voltage ferromagnetic Josephson junctions provide potential benefits as active elements in superconducting and quantum computing systems. Conventional metallic ferromagnetic Josephson junctions are typically over-damped, and are characterised by a significant generation of non-Cooper pair electrons, which are dissipative (so-called \textit{quasiparticle dissipation}). For applications in quantum computing logic, Josephson junctions require a high characteristic voltage and low damping, specifically a large subgap resistance to reduce quasiparticle tunnelling, which is detrimental for the coherence performance required for quantum computations~\cite{Kato2007,Serniak2018}. In Reference~\cite{Kawabata2006}, GdN has been proposed as the barrier layer for a ferromagnetic \textit{quiet} qubit, namely a flux-noise protected flux-qubit. In this application, the GdN-based Josephson junction must perform as a $\pi$ junction. In Reference~\cite{Ahmad2022b}, an alternative proposal for tunnel ferromagnetic Josephson junctions in superconducting qubits employs the transmon circuit design, namely a ferromagnetic transmon (ferro-transmon) qubit. In this work, the authors propose to integrate a tunnel ferromagnetic Josephson junction into a transmon to achieve an alternative tuning of the qubit frequency entirely relying on the hysteretic magnetic behaviour of the barrier, as in cryogenic memories applications. Further, it is in principle possible to operate these devices without static external magnetic fields, which are typically used to park the qubits at specific idling points for quantum computation, but at the same time are well-known be detrimental for coherence performances~\cite{siddiqi2021,Ahmad2023,Ahmad2025}. This is of course possible with any ferromagnetic material. However as discussed above the use of insulating ferromagnetic materials, like GdN, is motivated to reduce the effects of quasiparticle tunnelling. 

In Reference~\cite{Ahmad2022b} the authors also proposed a design to achieve a novel detector for the study of the intriguing physical phenomena that occur in hybrid superconducting/ferromagnetic systems. This proposal relies on the study of the coherence times of the device under different conditions, allowing one to highlight specific physical contributions to noise and fluctuations in the system. In order to do so, the ferro-transmon features a hybrid DC-SQUID, including a conventional Josephson junction and a tunnelling ferromagnetic Josephson junction, in order to achieve flexibility in the detection protocols. The possibility to include GdN-based devices featuring unconventional transport mechanisms, e.g., unconventional current-phase relations, $\pi$-states or spin-triplet superconductivity, has been envisioned, since this design may provide an alternative probe for the peculiar tunable, controllable magnetic behaviour of GdN barriers.

Motivated by this interest in forming ferromagnetic qubits based on GdN, a thorough analysis of the electro-dynamics of the NbN/GdN/NbN based Josephson junctions has been undertaken~\cite{Ahmad2020}. Here a Tunnel Junction Microscopic model was applied to these devices' current-voltage transport characteristics, to form a robust understanding of several key parameters required for their engineering. The results of this study pave the way for realistic circuit design. Reference~\cite{Ahmad2020} highlights the low frequency damping $Q_0$ in the previously studied NbN/GdN/NbN junctions, which is enhanced by up to two orders of magnitude compared to conventional metallic $\pi$-junctions. In fact, these GdN junctions have a comparable $Q_0$ to conventional AlO$_x$ based Josephson junctions commonly used in quantum circuits. These results imply that if functional $\pi$-junctions could be fabricated from GdN they would be appropriate for low dissipation, high coherence time \textit{quiet} qubits~\cite{Ahmad2020}, or alternatively as switchable novel hybrid superconducting qubits employing different circuit designs depending on the engineered Josephson and charging energy scales.

\section{Challenges, Outlook and Opportunities}

The fundamental understanding of the RN, and mixed cation (R,R')N alloys has seen significant progress in the previous decade, while the handful of demonstrations in superconducting electronics and resistive memory devices shows promise for this understanding to be leveraged in technological advances. Even so there are remaining challenges clear to the present authors. The most pressing being unequivocally determining the electronic ground state of the materials. 

One of the major themes of research in the RN in the modern era has been the manipulation of conductivity via the controlled inclusion of nitrogen vacancies. The majority of reports conclude that the ground state of the RN is insulating when no nitrogen vacancies are present. However, the direct measurement of the concentration of these vacancies has still not been accomplished. The clearest estimates are via two recent reports of epitaxial HoN~\cite{Pereira2023} and SmN~\cite{Melendez-Sans2024} thin films where the ratio of the nitrogen and lanthanide  XPS measurements is used to estimate the stoichiometry. However, these data have not yet been mapped clearly onto the resulting transport or structural properties of the films. In addition to the difficulties found in estimating the nitrogen vacancy concentration is the formation of well-ordered, near-stoichiometric thin films. The low activation energy of nitrogen vacancies during growth, near 1~eV~\cite{Punya2011,Porat2024}, implies a significant population of these will be present in films grown at the elevated temperatures required for epitaxial growth. These temperatures vary with lanthanide but are on the order of 200~$^\circ$C for HoN~\cite{Pereira2023}, 300-500$^\circ$C for SmN~\cite{Melendez-Sans2024,Anton2023,McNulty2021} and up to 800~$^\circ$C for GdN and DyN~\cite{natali2010,Anton2023}. 

Films grown at room temperature are polycrystalline, likely with significant contributions to the electronic properties from scattering events, and reports of localization behaviours~\cite{Holmes-Hewett2020}. Although much attention has been paid to nitrogen vacancies the other likely defects e.g., substitutional O, H, or various dislocations have yet been unexplored. Investigating defects beyond nitrogen vacancies may have significant impact, both addressing outstanding questions and for the potential for tuning the electronic properties through other means. A thorough study of the ground state of well ordered thin films, from stoichiometric to heavily doped, in which the nitrogen vacancy concentration is carefully measured, would serve well to settle debate regarding the insulating, or semi-metallic ground state of the RN. While the investigation of other likely defects would add breadth to the field. 

The electronic and magnetic properties of the RN materials have been the subject of study for decades. Even so, questions remain regarding the most studied members, including GdN and SmN. To date, the spin polarization of high-quality films has not been measured, though this is an important parameter for both fundamental understanding of their ground states as well as an indicator of their ultimate suitability for spin-dependent devices. 

The topic of the electronic and magnetic properties of alloys of (R,R')N has only been reported on very recently. These systems, described in detail in Section~\ref{S:Magnetic:Alloys}, perhaps harbour the greatest potential for technological applications. In this review, we have discussed the magnetization and angular momentum compensation points and their potential applications mostly in the context of work on (Gd,Sm)N. However, because an orbital-oriented magnetic moment is a characteristic found among the majority of light lanthanide ions, the investigation of other binary alloys combining light and heavy lanthanides, for example (Gd,Nd)N, should provide further material systems in which to realise these compensation points. This provides a fairly wide materials-space to explore these functionalities in devices such as those discussed in section~\ref{Applications}. In addition, the engineering of perpendicular magnetic anisotropy~\cite{Miller2023} or the exchange splitting of the conduction band~\cite{Porat2024} in these materials gives opportunities for useful application in Josephson devices, even for combinations of lanthanides in which the magnetic and angular momentum compensation are absent.

Unconventional electronic properties have at times been predicted in the RN, with recent articles suggesting the presence of topological states~\cite{Kim2018} and a Chern insulator state with potential for achieving the quantum anomalous Hall effect~\cite{Li2015}. Experimental evidence of unconventional states merits further investigation, such as the origin of the observed superconductivity in zero magnetic moment ferromagnet SmN. It has been suggested this is unconventional triplet superconductivity in films doped to near a quantum critical point~\cite{holmes-hewett2023}, and the superconducting state strengthens in bi- and multi-layers of SmN with strongly ferromagnetic GdN~\cite{Anton2016}. The exchange-spring behaviour of GdN/SmN~\cite{McNulty2019,Anton2021} may also be a route to generate spin-triplet superconducting currents by spin-mixing~\cite{banerjee2014}. 

Devices incorporating the RN have only just begun to be explored and there are many challenges left to be tackled if their unconventional magnetic properties are to be exploited in advanced electronics. For integration as memory or logic elements in cryogenic electronics, growth of high crystalline quality films on technologically-important superconducting materials such as Al, Nb or NbN is yet to be achieved. For coherent electronic transport, achieving smooth and clean interfaces between different materials is vital, as is avoiding interfacial reactions, such as those seen with high-temperature growth of RN on Si substrates~\cite{Natali2011}. Electrical control of the magnetisation state of individual elements is a requirement if RN memory arrays are to compete with the density of more conventional magnetic or semiconducting devices. Studies on spin-transfer or spin-orbit torque magnetisation dynamics and magnetisation switching of RNs and promising alloys are still outstanding. For the new sub-field of orbitronics, the presence of a large orbital magnetic moment in several RN has promise for generating large, long-ranged orbital torques~\cite{Ding2024,hayashi2023}. The ability to tailor the orbital and magnetic states in alloys of the (R,R')N presents an opportunity for materials with small magnetic moments and near zero angular momentum, but that generate and dissipate orbital torque highly efficiently. Using these materials it could be possible to achieve a long-standing goal of spintronics - magnetic memory devices that are nearly immune to magnetic fields. Such devices could be made to very high densities without cross-talk interference, and with a near zero angular momentum would be easily and rapidly switchable with small electrical currents.

\section{ACKNOWLEDGMENTS}

We would like to thank Joe Trodahl, Bob Buckley, Franck Natali and Avradeep Pal for their discussions and suggestions concerning this review, along with the many colleagues and students whose input has helped shape this document. H.~G.~A. acknowledges support of the PNRR MUR CN 00000013-ICSC and thanks the SUPERQUMAP project (COST Action CA21144). This research was supported by Quantum Technologies Aotearoa (Contract No. UOO2347), a research programme of Te Whai Ao—the Dodd Walls Centre, and the New Zealand Endeavour fund (Grant No. RSCHTRUSTVIC2447), both funded by the New Zealand Ministry of Business, Innovation and Employment. The MacDiarmid Institute is supported under the New Zealand Centres of Research Excellence Programme. The computations were performed on the R\={a}poi high performance computing facility of Victoria University of Wellington.

\bibliography{master.bib}

\end{document}